\documentclass[12pt]{article}
\usepackage{amsmath,amsthm,amssymb,epsfig,euscript,array,cancel,color}
\usepackage[colorlinks=true]{hyperref}
\usepackage{verbatim}
\usepackage{ulem}
\usepackage{url}
\usepackage{bbm}

\usepackage[paper=a4paper,dvips,top=2.54cm,left=2.74cm,right=2.34cm,
    foot=1cm,bottom=2.54cm]{geometry}

\theoremstyle{remark}

\newcounter{multieqs}

\newcommand{\be}{\begin{equation}}
\newcommand{\ee}{\end{equation}}
\newcommand{\eq}[1]{(\ref{#1})}

\newcommand{\bra}[1]{\langle #1|}
\newcommand{\ket}[1]{|#1 \rangle}
\newcommand{\ipr}[2]{\langle #1 | #2 \rangle}

\def\a{\alpha}        
\def\b{\beta}         
\def\g{\gamma}      
\def\d{\delta}      
\def\e{\epsilon}          
\def\f{\phi}          
\def\h{\eta}  

\def\l{{\lambda}}   
\def\m{\mu} \def\n{\nu}  
\def\o{\omega}  
\def\p{\pi}   
\def\r{\rho}  
\def\s{\sigma}    
  
\def\th{\theta} \def\Th{\Theta}   
  
\def\z{\zeta}

  \def\cF{{\cal F}}  
 \def\cH{{\cal H}}   
  \def\cL{{\cal L}}  
\def\cM{{\cal M}}    
\def\cP{{\cal P}}    
    
\def\cV{{\cal V}}

\def\Ft{\tilde{F}}
\def\Ht{\tilde{H}}

\def\pa{\partial}   
  
\def\uno{\mbox{1 \kern-.59em {\rm l}}}

\def\hlf{\frac{1}{2}}  
\def\ove#1{\frac{1}{#1}}

\def\bcomment#1{}  
  
\def\VEV#1{\left\langle #1\right\rangle}

\newcommand{\lrbrk}[1]{\left(#1\right)}

\def\w{{\wedge}}

\def\paP{\pa P}
\def\paPt{\widetilde{\pa P}}

\numberwithin{equation}{section}

\author{Pichet Vanichchapongjaroen
	\footnote{pichetv@nu.ac.th}
	\\
	\\
	{\small\it The Institute for Fundamental Study ``The Tah Poe Academia Institute'',}
	\\
	{\small\it Naresuan University, Phitsanulok 65000, Thailand}
}

\title{A Lagrangian relationship between the PST and the Sen formulations of chiral forms}

\begin{document}
\maketitle

\abstract{In this paper, we show the equivalence between the PST and Sen formulations for chiral forms
at the Lagrangian level. This is by discussing how an action in the Sen formulation for chiral forms
can be separated into two PST actions, one with the correct sign the other has the wrong sign,
reflecting the feature of the Sen formulation that it contains two chiral form fields, one physical the other unphysical, decoupled from each other.
The key idea is to add extra terms to the Sen action such that the modified action contains a PST scalar
and is also equivalent with the original one. Then after eliminating a self-dual field $Q$ and making appropriate field redefinitions, we obtain a sum of two PST actions.
We also consider alternative actions in which extra terms are added to the Sen action
to either eliminate the physical or unphysical degrees of freedom
and such that the resulting action is identified with the PST action
either with correct or incorrect sign. Generalisation to non-linear theories coupling to external source is discussed.
The realisation of the Sen M5-brane action as containing the PST M5-brane action
is also explicitly shown.}

\thispagestyle{empty}
\newpage
\tableofcontents

\section{Introduction}

Chiral form fields have important applications in string theory and M-theory.
For example, a chiral $0$-form in $2$ dimensions can be used
to explain a heterotic string \cite{Gross:1984dd}.
A chiral $2$-form in $6$ dimensions is a field on an M5-brane \cite{Gueven:1992hh}.
A chiral $4$-form in $10$ dimensions is a field in the type IIB string theory \cite{Green:1981yb}.

A chiral $2p$-form $A$ in
a $(4p+2)$-dimensional spacetime
with metric $g$
is a gauge field whose field strength $F = dA$
is self-dual or anti-self-dual.
At the linear level, the linear (anti-)self-duality condition reads
$
*F = \pm F,
$
where the plus sign ($+$) is for self-duality,
while the minus sign ($-$) is for anti-self-duality.
Let us call the chiral form whose field strength is self-dual (resp. anti-self-dual) as
being $(+1)$-handed (resp. $(-1)$-handed).

A naive attempt to write down a Lagrangian for
a chiral form
whose field strength is linear (anti-)self-dual
is $F\w *F.$ However, this Lagrangian simply vanishes
upon imposing (anti-)self-duality condition.

Various attempts have been made in the literature
which provide important insights to solve the problem
but are faced with some shortcomings.
The idea which is useful is that the (anti-)self-duality condition
should be introduced into the 
Maxwell $2p$-form theory (whose Lagrangian is of the form $\sim F\w*F$)
in a $4p+2$ dimensional spacetime.
As an example, in \cite{Siegel:1983es}
squares of second-class constraints are introduced,
making them incorrectly identified with first-class constraints.
Another example is by \cite{McClain:1990sx}.
The (anti-)self-duality condition is imposed by second-class constraints.
The process of turning all the constraints into the first-class ones
takes infinitely many steps.

A successful attempt \cite{Floreanini:1987as}, \cite{Henneaux:1988gg}, to be dubbed ``HT formulation'',
is carried out with the help of the Hamiltonian analysis
which essentially splits the spacetime coordinates $x^\m; \m \in \{0,1,\cdots,4p+1\}$
into time and spatial coordinates as $x^0, x^i$; $i\in\{1,\cdots,4p+1\}$.
The resulting action is still diffeomorphism invariance
but is non-manifestly with non-standard diffeomorphism transformation rule.
A related successful attempt \cite{Perry:1996mk}, to be dubbed ``PS formulation'',
is carried out based on the feature that in the HT formulation
the spacetime indices are split.
In the PS formulation, however, a different splitting on the spacetime coordinates
is made.
In this case, the splitting gives $x^{\hat{\m}}, x^{4p+1}$; $\hat{\m}\in\{0,1,\cdots,4p\}$.
A generalisation is also carried out \cite{Chen:2010jgb}, \cite{Huang:2011np}
by making various alternatives for the splitting of spacetime coordinates.
Extensions of the quadratic actions for chiral $2$-forms
in the formulations with non-manifest diffeomorphism invariance
to describe a single M5-brane
can also be achieved \cite{Aganagic:1997zq}, \cite{Ko:2016cpw}.

In order to restore the manifest diffeomorphism invariance,
an auxiliary scalar field $a$ is introduced.
This approach is the PST formulation \cite{Pasti:1995ii}, \cite{Pasti:1996vs}, \cite{Pasti:1997gx}, \cite{Bandos:1997ui}.
Generalisations to more than one auxiliary scalar field
as well as extensions to single M5-brane actions
is possible \cite{Pasti:2009xc}, \cite{Ko:2013dka}, \cite{Ko:2017tgo}.

There is an alternative formulation for describing chiral form.
This formulation the Sen formulation \cite{Sen:2015nph}, \cite{Sen:2019qit},
which is inspired from the Sen formulation of string field theory \cite{Sen:2015uaa}.
This formulation is further developed in \cite{Andriolo:2020ykk}, \cite{Hull:2023dgp}.
In the recent version, the Sen formulation describes
two $(+1)$-handed $2p$-forms in a $4p+2$ dimensional spacetime.
One of the chiral forms is chiral with respect to a metric $g$,
while the other is chiral with respect to another metric $\bar{g}$.
Let us call them as $g$-chiral form and $\bar{g}$-chiral form, respectively.
One of them is physical while the other is unphysical due to incorrect sign of kinetic term.
Following the convention of \cite{Hull:2023dgp},
the $g$-chiral form is physical
while the $\bar{g}$-chiral form is unphysical.
These two chiral forms are uncoupled from each other.

The key features for the set up of the Lagrangian
description of the Sen formulation are as follows.
The independent fields in the Sen formulation are
a $2p$-form field $P$ and a $(2p+1)$-form field $Q$,
which is linearly self-dual with respect to $\bar{g}$.
A certain combination $H$ of $Q, g, \bar{g}$
is self-dual with respect to $g$. The field $H$ should be closed on-shell.

Further insights on the Sen formulation are being developed
for example in \cite{Vanichchapongjaroen:2020wza}, \cite{Andriolo:2021gen}, \cite{Evnin:2022kqn}.

The description in the two formulations are different in nature.
In particular, in a topologically trivial spacetime,
while a $(2p+1)$-form field in the PST formulation
is exact off-shell and is self-dual on-shell,
a $(2p+1)$-form field in the Sen formulation
is self-dual off-shell and is exact on-shell.
The interchange of the role of a $(2p+1)$-form field
make it a priori unclear whether the two formulations can be related
at the Lagrangian level.

The literature suggests that this could be possible.
The works \cite{Andriolo:2020ykk}, \cite{Phonchantuek:2023iao}
suggest that the double dimensional reduction can be performed
on the Sen M5-brane action \cite{Vanichchapongjaroen:2020wza}
giving rise to the D4-brane action,
which also obtainable from a dimensional reduction of the PST M5-brane action \cite{Pasti:1997gx}, \cite{Bandos:1997ui}. Key ideas from the double dimensional reduction
of the Sen M5-brane action are that one eliminates the field $Q$,
and makes use of appropriate field redefinition.

The relationship between the Sen formulation
and the HT formulation
is realised in \cite{Janaun:2024wya}.
Essentially, the ideas explained in the previous paragraph are made use.
It turns out however that in contrast to the analysis of the double dimensional reduction
which works entirely using Lagrangian,
the realisation of the relationship 
between the Sen formulation
and the HT formulation requires
transforming to the phase space.
The field redefinition is also made on the phase space.
This procedure seems to be only suitable to obtaining the HT formulation
as the common feature between the Hamiltonian analysis and the HT formulation
is that time and space coordinates are treated differently.
Due to this reason, it is not clear how to obtain the PST formulation form the Sen formulation
by following similar approach.
In particular, it is not clear how the auxiliary PST scalar $a$ should enter during the process.

It should be more preferable to look for a different approach,
which works entirely at the Lagrangian level.
The main goal of this paper is to realise
the relationship between the Sen formulation
and the PST formulation.

This paper is organised as follows.
In section \ref{sec:review},
we review the PST and the Sen formulations
at the quadratic level.
We also review how to realise the Sen action
as a sum of two HT actions
and explain how it is difficult to follow the same procedure
to obtain two PST actions from the Sen action.
In section \ref{sec:Sen-separate},
we first discuss a toy model which realises
a Maxwell $2p$-form action
as a sum of two PST actions.
Then we apply similar idea to the quadratic Sen action
as well as making consistency checks along the way.
In section \ref{sec:Sen-constraint},
we introduce extra terms to the quadratic Sen action
such that, depending on the extra terms,
either the $\bar{g}$-chiral or the $g$-chiral
is eliminated from the theory.
In section \ref{sec:Sen-nl},
we extend the idea to non-linear Sen actions
with $(2p+1)$-form source.
We also demonstrate with explicit calculation
that the Sen M5-brane action
can be separated into a sum of the PST M5-brane action
and a quadratic PST action of uncoupled unphysical chiral $2p$-form.
Finally in section \ref{sec:conclusion}, we conclude, discuss, as well as suggesting future works.

\textbf{Convention: }
Coordinates of the $(4p+2)$-dimensional spacetime are denoted by $x^\m,$
with $\m\in\{0,1,\cdots,4p+1\}$.
Other Greek alphabet, for example, $\a,\b,\g,\n,\r$ may also be used as spacetime indices.
The time coordinate is $t\equiv x^0$.
The spatial coordinates are denoted $x^i,$
with $i\in\{1,2,\cdots,4p+1\}.$
Other lower-case Roman alphabet, for example, $a,b,c,j,k,m,n,p$ may also be used as space indices.
Differential $q$-forms $\o_{(q)}$ are expressed in coordinate basis as
\be
\o_{(q)}
\equiv
\ove{q!}\o_{\m_1\cdots \m_q}dx^{\m_q}\w\cdots \w dx^{\m_1}.
\ee 
Exterior derivatives and interior products
act from the right. For example, $d(\a_{(q)}\w\b_{(r)}) = \a_{(q)}\w d\b_{(r)} + (-1)^r d\a_{(q)}\w\b_{(r)},$ for any $q$-form $\a_{(q)}$ and $r$-form $\b_{(r)}$.
There are two Hodge star operators $*, \bar{*}$ corresponding to the two metric $g, \bar{g}.$
In particular,
\be
*(dx^{\m_1}\w\cdots\w dx^{\m_{2p+1}})
=\frac{\sqrt{-\det{g}}}{(2p+1)!}dx^{\n_1}\w\cdots\w dx^{\n_{2p+1}}g_{\n_1\r_1}\cdots g_{\n_{2p+1}\r_{2p+1}}\e^{\r_1\cdots\r_{2p+1}\m_1\cdots\m_{2p+1}},
\ee
\be
\bar{*}(dx^{\m_1}\w\cdots\w dx^{\m_{2p+1}})
=\frac{\sqrt{-\det{\bar{g}}}}{(2p+1)!}dx^{\n_1}\w\cdots\w dx^{\n_{2p+1}}\bar{g}_{\n_1\r_1}\cdots \bar{g}_{\n_{2p+1}\r_{2p+1}}\e^{\r_1\cdots\r_{2p+1}\m_1\cdots\m_{2p+1}},
\ee
where $\e^{\r_1\cdots\r_{2p+1}\m_1\cdots\m_{2p+1}}$ is the Levi-Civita symbol
defined such that $\e^{01\cdots(4p+1)} = 1$.
We also define the Levi-Civita symbol with only the spatial indices as $\e^{i_1\cdots i_{4p+1}}=\e^{0i_1\cdots i_{4p+1}}$.

\section{A review of the PST and the Sen formulation at quadratic level}\label{sec:review}
\subsection{The PST action}\label{subsec:PST}

The quadratic PST action for a $(\pm 1)$-handed chiral $2p$-form $A$
(with field strength $F = dA$)
in a $4p+2$ dimensional spacetime with metric $g$ is given by \cite{Pasti:1995ii}, \cite{Pasti:1996vs}, \cite{Pasti:1996va}
\be\label{PST-0}
S_{\textrm{PST}}^{(g), (\pm 1)} = \int\cL_{\textrm{PST}}^{(g), (\pm 1)},
\ee
with
\be\label{PST-action}
\begin{split}
	\cL_{\textrm{PST}}^{(g), (\pm 1)}
	&=-\ove{4}F\w*F \pm \frac{1}{4}\cF_{\pm}\w\cP\cF_{\pm},
\end{split}
\ee
where
\be\label{defncP}
\cP
\equiv\frac{da\w i_{g^{-1}da}}{(\pa a)^2},
\ee
\be
\cF_{\pm}
\equiv
F \mp *F,
\ee
\be
g^{-1}da
\equiv g^{\m\n}\pa_\m a\pa_\n,
\ee
\be
(\pa a)^2
\equiv g^{\m\n}\pa_\m a\pa_\n a.
\ee
The field $a$ is an auxiliary scalar field
with no dynamics.

In order to analyse the theory \eq{PST-0},
one first considers the variation with respect to $A$ and $a$.
It is convenient to use the identities
\be
\d_a(\cF_{\pm}\w\cP\cF_{\pm})
=-2d\d a\w\frac{i_{g^{-1}da}\cF_{\pm}}{(\pa a)^2}\w\cP\cF_{\pm},
\ee
\be
\m\w*\l = \l\w*\m,
\ee
\be
\m\w\cP \l = \l\w(\cP-1)\m,
\ee
\be
*\cP\l = (1-\cP)*\l,
\ee
\be
*^2\l = \l,
\ee
for any $(2p+1)$-forms $\m$ and $\l$.
The variation can then easily be obtained as
\be\label{vary-PST}
\d\cL_{\textrm{PST}}^{(g,\pm 1)}
=\pm\lrbrk{d\d A-\frac{1}{2} d\d a\w\frac{i_{g^{-1}da}\cF_{\pm}}{(\pa a)^2}}\w\cP\cF_{\pm}
+tot.,
\ee
where $tot.$ is a collection of total derivative terms.
From the variation, it is easy to see that
apart from the tensor gauge symmetry in $A,$ i.e. $A\to A+d\l$ for any $(2p-1)$-form $\l$,
there are two important symmetries called, following \cite{Bandos:2014bva}, PST1 and PST2 symmetries.
The PST1 symmetry is
\be\label{PST1-PST}
\d A = -\f da,\qquad
\d a = 0,
\ee
where $\f$ is an arbitrary $(2p-1)$-form.
This symmetry is a gauge symmetry
used to transform the solution of the equation of motion (EOM)
to self-duality condition.\footnote{In the case $p=0$, the PST1 symmetry eq.\eq{PST1-PST}
is not valid as $\f$ would be a $(-1)$-form. A different treatment is required. See for example \cite{Bandos:2014bva},
in which a semi-local symmetry $\d A(t,x) = f(a(t,x)), \d a = 0$ is being used instead.
Only when $da$ is timelike, the semi-local symmetry can be used
to show the equivalence between the equation of motion and the self-duality condition.
However, when $da$ is spacelike, the equation of motion is not equivalent to the self-duality condition.
On the contrary, in the case $p>0$, the PST theory is chiral
in both $da$-timelike case and $da$-spacelike case.}${}^{,}$\footnote{Despite the nuance differences, as discussed in the previous footnote, between the cases 
$p = 0$ and $p>0$, we will simply mention these two cases collectively for brevity.}
Next, the PST2 symmetry is
\be\label{PST2-PST}
\d A = \d a \frac{i_{g^{-1}da}\cF_{\pm}}{(\pa a)^2},
\ee
where $\d a$ is arbitrary.
This symmetry is a gauge symmetry used to gauge away the auxiliary field $a$.

Let us now consider equations of motion for $A$ and $a$.
From the variation \eq{vary-PST},
one obtains
\be\label{EOMP-PST}
d(\cP\cF_{\pm}) = 0,
\ee
\be\label{EOMa-PST}
d(\cP\cF_{\pm})\w\frac{i_{g^{-1}da}\cF_{\pm}}{(\pa a)^2} = 0.
\ee
From the equations of motion, it is immediately clear that
the field $a$ has no dynamics
since its equation of motion is implied by the equation of motion of $A$.
Next, from eq.\eq{EOMP-PST},
the general solution is \cite{Pasti:1996vs},
\be\label{presolnP-PST}
\cP\cF_{\pm}
=d\th\w da,
\ee
where $\th$ is an arbitrary $(2p-1)$-form.
Under the PST1 transformation 
\eq{PST1-PST}
with parameter $\f = -\th$,
the equation \eq{presolnP-PST}
transforms to
\be
\cP\cF_{\pm} = 0,
\ee
which after using (anti-)self-duality of $\cF_{\pm},$
one obtains
\be
\cF_{\pm} = 0,
\ee
or
\be
F = \pm*F,
\ee
which is the required (anti-)self-duality condition for $F.$

One may recover HT and PS actions by appropriately
fixing the PST2 gauge symmetry of the PST action.
In particular, to obtain the HT action \cite{Henneaux:1988gg},
one may fix the gauge $a = t$
on the theory \eq{PST-action}.
The resulting action is simply
\be\label{HT-action}
\begin{split}
	\cL_{\textrm{HT}}^{(g), (\pm 1)}
	&=\cL_{\textrm{PST}}^{(g), (\pm 1)}\bigg|_{a = t}\\
	&=-\ove{4}dA\w*dA \pm \frac{1}{4}\cF_{\pm}\w\cP_t\cF_{\pm},
\end{split}
\ee
where
\be
\cP_t
\equiv\frac{da\w i_{g^{-1}dt}}{(\pa t)^2}.
\ee
The HT action \eq{HT-action} in invariance under the modified diffeomorphism $\d_\xi$,
which is the sum
of the PST2 transformation $\d_{PST2}$
and the standard diffeomorphism $\pounds_\xi$,
where we denote Lie derivative using the notation $\pounds$.
This means that the modified diffeomorphism is given by
\be
\d_\xi = \d_{PST2} + \pounds_\xi.
\ee
It is also required that the modified diffeomorphism
leaves $a$ in the gauge $a = t$, i.e. $\d_\xi a = 0$.
This implies that
\be
\d_{PST2}a = -\xi^0,
\ee
which fixes the parameter for the PST2 transformation.
The modified diffeomorphism on $A$ is then
\be\label{mod-diff-HT}
\d_{\xi}A
=\pounds_\xi A
-\frac{\xi^0 i_{g^{-1}dt}\cF_{\pm}}{g^{00}},
\ee
which agrees with \cite{Henneaux:1988gg}.
As an explicit check, note that
\be
\pounds_\xi\cL_{\textrm{PST}}^{(g), (\pm 1)} = tot.
\ee
In order for the modified diffeomorphism
to be a symmetry of the HT action,
one requires
\be
\d_\xi\cL_{\textrm{HT}}^{(g), (\pm 1)} = tot.
\ee
Therefore,
\be
\d_\xi\cL_{\textrm{HT}}^{(g), (\pm 1)}
=tot. + \lrbrk{\pounds_\xi\cL_{\textrm{PST}}^{(g), (\pm 1)}}\bigg|_{a = t}.
\ee
By reading off from eq.\eq{vary-PST} and noting that
the diffeomorphism rule on $g$ is not modified,
one obtains
\be
d\d_\xi A\w\cP_t\cF_{\pm}
=
\lrbrk{d\pounds_\xi A-\frac{1}{2} d\pounds_\xi t\w\frac{i_{g^{-1}dt}\cF_{\pm}}{g^{00}}}\w\cP_t\cF_{\pm}
+tot.,
\ee
which implies
\be
\d_\xi A\w d\cP_t\cF_{\pm}
=\lrbrk{\pounds_\xi A 
-i_\xi dt\ \frac{i_{g^{-1}dt}\cF_{\pm}}{g^{00}}}\w d\cP_t\cF_{\pm}
+tot.,
\ee
which is satisfied by eq.\eq{mod-diff-HT}.

Similarly, one may obtain the PS action
by fixing the gauge $a = x^{4p+1}$
on the PST action \eq{PST-0}.
The upshot is that
\be
	\cL_{\textrm{PS}}^{(g), (\pm 1)}
	=\cL_{\textrm{PST}}^{(g), (\pm 1)}\bigg|_{a = x^{4p+1}}.
\ee

\subsection{The Sen action}

The independent fields in the Sen theory are 
a $2p$-form $P$ and a $(2p+1)$-form field $Q$ which is self-dual with respect to metric $\bar{g}.$ That is $Q = \bar{*}Q,$
where $\bar{*}$ is the Hodge star operator with respect to $\bar{g}$.
A certain combination $H$ of $Q, g, \bar{g}$ is self-dual with respect to $g$.
For example, in the quadratic Sen theory, 
$H = *H$.

In more details, the
Lagrangian for the quadratic Sen theory is given by
\be\label{quadSen}
\begin{split}
	\cL_{\textrm{Sen}} =
	\ove{4}dP\w\bar{*} dP - Q\w dP +\ove{2}Q\w R,
\end{split}
\ee
where $Q$ satisfies
\be\label{Qsds}
Q = \bar{*}Q,
\ee
and
$R$ is defined through
\be
\d_Q(Q\w R)
=2\d Q\w R.
\ee
It also satisfies
\be\label{Rantisd}
R  = -\bar{*}R,
\ee
and
\be\label{Hsd}
H = *H,
\ee
where
\be
H = Q - R.
\ee
The conditions \eq{Rantisd}-\eq{Hsd} completely fix $R.$
It is convenient to write
\be\label{RMQ}
R = \cM Q,
\ee
where $\cM$ is a map from a $\bar{*}$-self-dual $(2p+1)$-form
to a $\bar{*}$-anti-self-dual $(2p+1)$-form
and satisfies
\be\label{propM-1}
\lrbrk{\frac{1-\bar{*}}{2}}\cM\lrbrk{\frac{1+\bar{*}}{2}}
=\cM\lrbrk{\frac{1+\bar{*}}{2}},
\ee
\be\label{propM-2}
\lrbrk{\frac{1+*}{2}}(1-\cM)\lrbrk{\frac{1+\bar{*}}{2}}
=(1-\cM)\lrbrk{\frac{1+\bar{*}}{2}},
\ee
\be\label{propM-3}
\int \m\w\cM \l
=\int \l\w\cM \m,
\ee
where $\m$ and $\l$ are any $\bar{*}$-self-dual $(2p+1)$-form.
In order to evaluate $\cM,$
one may extend the domain of $\cM$
by defining how it maps a $\bar{*}$-anti-self-dual $(2p+1)$-form.
Various definitions are presented in the literature \cite{Sen:2015nph}, \cite{Andriolo:2020ykk}, \cite{Vanichchapongjaroen:2020wza}.
However, the explicit form of $\cM$ is not needed in this paper.

The two chiral forms in the theory \eq{quadSen}
are decoupled.
This can be shown \cite{Sen:2019qit}, \cite{Andriolo:2020ykk}, \cite{Evnin:2022kqn}, \cite{Janaun:2024wya}
at the Hamiltonian level, equation of motion level,
as well as Lagrangian level.
In particular, let us demonstrate this at the equation of motion level.
The equations of motion for $P$ and $Q$ of the Lagrangian \eq{quadSen}
is
\be\label{EOMP-Sen}
d\lrbrk{\ove{2}\bar{*}dP + Q}
=0.
\ee
The equation of motion of $Q$ is
\be\label{EOMQ-Sen}
-\lrbrk{\frac{1-\bar{*}}{2}}dP + R
=0.
\ee
Denoting
\be\label{Hsdefn}
H_{(s)}(Q,P)
\equiv
Q + \ove{2}(1+\bar{*})dP,
\ee
the general solution to eq.\eq{EOMP-Sen}
is
\be\label{soln1-Sen}
H_{(s)}(Q,P)
=
-\hlf d\hat{P} + \hlf dP,
\ee
where $\hat{P}$ is an arbitrary $2p$-form.
Since $H_{(s)}$ is $\bar{*}$-self-dual,
we see that $-\hat{P}/2 + P/2$ is $(+1)$-handed $\bar{g}$-chiral.
Next, combining eq.\eq{EOMQ-Sen} and eq.\eq{soln1-Sen}
gives
\be\label{soln2-Sen}
\begin{split}
	H
	&=-\hlf d\hat{P} - \hlf dP.
\end{split}
\ee
Since $H$ is $*$-self-dual,
we see that $-\hat{P}/2 - P/2$ is $(+1)$-handed $g$-chiral.

\subsection{A separation of the Sen action to two HT actions}
At the Hamiltonian and Lagrangian levels \cite{Sen:2019qit}, \cite{Andriolo:2020ykk}, \cite{Janaun:2024wya},
the decoupling between the $\bar{g}$-chiral and the $g$-chiral fields
are also realised. Furthermore,
the $\bar{g}$-chiral field is unphysical
since its kinetic term has the wrong sign.
As for the $g$-chiral field,
its kinetic term has the correct sign,
so it is physical.

In particular, in \cite{Janaun:2024wya}, based on constructions and insights from \cite{Sen:2019qit}, \cite{Andriolo:2020ykk}, \cite{Tseytlin:1990ar}, \cite{Sonnenschein:1988ug},
the Sen action can be expressed as two HT actions.
To quickly review this and to discuss why the method should be modified
when generalising from HT case to PST case,
let us focus on the flat case $g = \bar{g} = \h$.
The Lagrangian \eq{quadSen} can be written in index notation as, with $d^{4p+2}x/((2p)!)$ omitted,
\be
\cL_{\textrm{Sen}}
=-\ove{4}\paP_{a_1\cdots a_{2p}}\paP^{a_1\cdots a_{2p}}
+\ove{4}\paPt_{a_1\cdots a_{2p}}\paPt^{a_1\cdots a_{2p}}
+Q_{a_1\cdots a_{2p}}(\paPt^{a_1\cdots a_{2p}}
-\paP^{a_1\cdots a_{2p}}),
\ee
where
\be
\paP_{a_1\cdots a_{2p}}
=(2p+1)\pa_{[0}P_{a_1\cdots a_{2p}},
\ee
\be
\paPt^{a_1\cdots a_{2p}}
=-\frac{1}{(2p)!}\pa_{[i}P_{b_1\cdots b_{2p}]}\e^{a_1\cdots a_{2p}ib_1\cdots b_{2p}},
\ee
and
\be
Q_{a_1\cdots a_{2p}}\equiv Q_{0a_1\cdots a_{2p}}.
\ee
Due to self-duality of $Q$, only half of the components are present in the Lagrangian.
Furthermore, in the case $g = \bar{g} = \h$, we have $R = 0.$
Conjugate momenta for $P_{\m_1\cdots\m_{2p}}$ and $Q_{a_1\cdots a_{2p}}$
are denoted $\p^{\m_1\cdots\m_{2p}}$ and $\p_Q^{a_1\cdots a_{2p}}$.
The total Hamiltonian is
\be\label{cH-1}
\begin{split}
	\cH_T
	&=-(\p^{a_1\cdots a_{2p}} + Q^{a_1\cdots a_{2p}})(\p_{a_1\cdots a_{2p}} + Q_{a_1\cdots a_{2p}})
	-\ove{4}\paPt_{a_1\cdots a_{2p}}\paPt^{a_1\cdots a_{2p}}\\
	&\qquad-Q_{a_1\cdots a_{2p}}\paPt^{a_1\cdots a_{2p}}
	 - 2p\p^{a_1 a_2\cdots a_{2p}}\pa_{a_1}P_{a_2\cdots a_{2p}0}\\
	&\qquad+\z_{a_1\cdots a_{2p-1}}\p^{0 a_1\cdots a_{2p-1}}
	+u_{a_1\cdots a_{2p}}\p_Q^{a_1\cdots a_{2p}},
\end{split}
\ee
where $\z_{a_1\cdots a_{2p-1}}$ and $u_{a_1\cdots a_{2p}}$ are Lagrange multipliers
which enforce the primary constraints $\p^{0 a_1\cdots a_{2p-1}}\approx 0$ and $\p_Q^{a_1\cdots a_{2p}}\approx 0.$
In this theory, the first class constraints are
$\p^{0 a_1\cdots a_{2p-1}}\approx 0$ and $\pa_{a_1}\p^{a_1\cdots a_{2p}}\approx 0,$
whereas second-class constraints are
$\p_Q^{a_1\cdots a_{2p}}\approx 0$
and $Q^{a_1\cdots a_{2p}} + \p_+^{a_1\cdots a_{2p}} \approx 0,$
where
\be
\p_\pm^{a_1\cdots a_{2p}}
=\p^{a_1\cdots a_{2p}} \pm \hlf\paPt^{a_1\cdots a_{2p}}.
\ee
Solving the second-class constraints for
$\p_Q^{a_1\cdots a_{2p}}$ and $Q_{a_1\cdots a_{2p}}$
gives
\be\label{piQsoln-2ndclass}
\p_Q^{a_1\cdots a_{2p}} \approx 0,
\ee
\be\label{Qsoln-2ndclass}
Q_{a_1\cdots a_{2p}}
\approx-(\p_+)_{a_1\cdots a_{2p}}.
\ee
Substituting eq.\eq{piQsoln-2ndclass}-\eq{Qsoln-2ndclass}
into eq.\eq{cH-1} gives
\be\label{HT_sep}
\begin{split}
\cH_T
&=-\hlf\p_-^{a_1\cdots a_{2p}}(\p_-)_{a_1\cdots a_{2p}}\\
&\qquad
+\hlf\p_+^{a_1\cdots a_{2p}}(\p_+)_{a_1\cdots a_{2p}}\\
&\qquad+ 2p\pa_{a_1}\p^{a_1 a_2\cdots a_{2p}}P_{a_2\cdots a_{2p}0}
+\z_{a_1\cdots a_{2p-1}}\p^{0 a_1\cdots a_{2p-1}}.
\end{split}
\ee
The first-order Lagrangian corresponding to the total Hamiltonian \eq{HT_sep}
is given by
\be
\cL_T
=\p^{a_1\cdots a_{2p}}\pa_0 P_{a_1\cdots a_{2p}}
+2p\p^{0a_1\cdots a_{2p-1}}\pa_0 P_{0a_1\cdots a_{2p-1}}
-\cH_T,
\ee
which becomes, after
eliminating $\p^{0a_2\cdots a_{2p}}$ and $\z_{a_2\cdots a_{2p}}$ using their equations of motion,
\be
\begin{split}
\cL_T
&=\p^{a_1\cdots a_{2p}}\pa_0 P_{a_1\cdots a_{2p}}
+\hlf\p_-^{a_1\cdots a_{2p}}(\p_-)_{a_1\cdots a_{2p}}\\
&\qquad
-\hlf\p_+^{a_1\cdots a_{2p}}(\p_+)_{a_1\cdots a_{2p}}\\
&\qquad- 2p\pa_{a_1}\p^{a_1 a_2\cdots a_{2p}}P_{a_2\cdots a_{2p}0}.
\end{split}
\ee
By using the field redefinitions
\be
P_{a_1\cdots a_{2p}}
=\r_{a_1\cdots a_{2p}} + \s_{a_1\cdots a_{2p}},
\ee
\be
\p^{a_1\cdots a_{2p}}
=\hlf\widetilde{\pa\r}^{a_1\cdots a_{2p}}
-\hlf\widetilde{\pa\s}^{a_1\cdots a_{2p}},
\ee
where
\be
\widetilde{\pa\r}^{a_1\cdots a_{2p}}
\equiv-\frac{1}{(2p)!}\pa_{[i}\r_{b_1\cdots b_{2p}]}\e^{a_1\cdots a_{2p}ib_1\cdots b_{2p}},
\ee
\be
\widetilde{\pa\s}^{a_1\cdots a_{2p}}
\equiv-\frac{1}{(2p)!}\pa_{[i}\s_{b_1\cdots b_{2p}]}\e^{a_1\cdots a_{2p}ib_1\cdots b_{2p}},
\ee
one obtains
\be\label{LT}
\begin{split}
	\cL_T
	&=-\hlf\widetilde{\pa\s}^{a_1\cdots a_{2p}}\dot\s_{a_1\cdots a_{2p}}
	+\hlf\widetilde{\pa\s}^{a_1\cdots a_{2p}}\widetilde{\pa\s}_{a_1\cdots a_{2p}}\\
	&\qquad
	+
	\hlf\widetilde{\pa\r}^{a_1\cdots a_{2p}}\dot\r_{a_1\cdots a_{2p}}
	-\hlf\widetilde{\pa\r}^{a_1\cdots a_{2p}}\widetilde{\pa\r}_{a_1\cdots a_{2p}},
\end{split}
\ee
which describes the unphysical chiral $(2p)-$form $\s$
uncoupled with the physical chiral $(2p)-$form $\r.$
Each part of the Lagrangian \eq{LT}
is in the form of the HT Lagrangian
but such that $\s$ has the wrong sign,
whereas $\r$ has the correct sign.

The procedure above seems to only be suitable to obtain the HT Lagrangians.
It would be more difficult if at all possible
to obtain the PS Lagrangians, which splits one spatial direction
instead of splitting the time direction as in the case of the HT Lagrangians.
It is expectedly even more difficult to obtain the PST Lagrangians.
For example, the only place to introduce the PST scalar $a$
seems to be at the field redefinition.
However, after some initial calculations we expect that this is not likely possible.

A better approach is to work entirely at the Lagrangian level.
The idea is to consider a Lagrangian equivalent to the original one
but with PST scalar $a$ introduced. This is to be discussed in the next section.

\section{Separations of the quadratic Sen action}\label{sec:Sen-separate}
\subsection{Motivation: Maxwell $2p$-form action as a sum of two PST actions}

Consider a Maxwell $2p$-form Lagrangian
\be
\cL_{\textrm{Maxwell}} = -\ove{4}F\w*F,
\ee
where $F = dA$
is the $(2p+1)$-form field strength of a $2p$-form
gauge field $A$.
The equation of motion
\be
d*F = 0
\ee
is, on spacetime with trivial topology, equivalent to
\be\label{FTheta}
F = *\Th,
\ee
where $\Th = d\th$
is the $(2p+1)$-form field strength of a $2p$-form gauge field $\th$.

The Lagrangian which gives rise to the equation of motion \eq{FTheta}
is
\be\label{L_Ftheta}
\begin{split}
\tilde{\cL}_{\textrm{Maxwell}}
&= -\ove{4}F\w*F - \ove{4}Y\w\cP*Y\\
&= \hlf\Th\w\cP F - \ove{4}\Th\w*(1-\cP)\Th - \ove{4}F\w*(1-\cP)F,
\end{split}
\ee
where $Y = \Th - *F, F = dA, \Th = d\th$,
and $\cP$ is as defined in eq.\eq{defncP}.
The variation of the Lagrangian \eq{L_Ftheta}
is
\be
\begin{split}
\d\tilde{\cL}_{\textrm{Maxwell}}
&=-\hlf\cP Y\w\d F
+\hlf\cP *Y\w \d\Th
+\hlf d\d a\w\frac{\cP*X}{(\pa a)^2}\w i_{g^{-1}da}Y\\
&=tot. + \hlf\lrbrk{\d A-\d a\frac{i_{g^{-1}da}*Y}{(\pa a)^2}}\w d\cP Y
-\hlf\lrbrk{\d \th+\d a\frac{i_{g^{-1}da}Y}{(\pa a)^2}}\w d\cP *Y.
\end{split}
\ee
From the variation, it can be seen that
apart from the tensor gauge transformations for $A$ and $\th$,
there are also symmetries called PST1 and PST2 symmetries.
The PST1 symmetry is
\be
\d A = \chi^{(1)}\w da,\qquad
\d\th = \chi^{(2)}\w da.
\ee
It will be used
in order to transform
the solution to the equation of motion to eq.\eq{FTheta}.
The PST2 symmetry is
\be
\d A
=\d a\frac{i_{g^{-1}da}*Y}{(\pa a)^2},\qquad
\d\th
=-\d a\frac{i_{g^{-1}da}Y}{(\pa a)^2}.
\ee
The PST2 symmetry is related to the auxiliary nature of $a$.

EOM of $A, \th, a$ are, respectively,
\be\label{M_EOMA}
d(\cP Y) = 0,
\ee
\be\label{M_EOMth}
d(\cP *Y) = 0,
\ee
\be
\lrbrk{\frac{i_{g^{-1}da}*Y}{(\pa a)^2}}\w d(\cP Y)
+\lrbrk{\frac{i_{g^{-1}da}Y}{(\pa a)^2}}\w d(\cP*Y)
=0.
\ee
It is clear that EOM of $a$
follows from other EOM.
We are then left with
eq.\eq{M_EOMA}-\eq{M_EOMth}
which are equivalent to
\be
\cP Y = d\psi^{(1)}\w da,\qquad
\cP*Y = d\psi^{(2)}\w da.
\ee
By making use of PST1 gauge transformations
with gauge parameters $\chi^{(1)} = -\psi^{(2)}, \chi^{(2)} = \psi^{(1)},$
we obtain
\be
\cP Y = 0,\qquad
\cP*Y = 0,
\ee
which is equivalent to
\be
Y = 0,
\ee
i.e. eq.\eq{FTheta}.

After a field redefinition,
the Lagrangian \eq{L_Ftheta}
can be written as a sum
of two PST Lagrangians.
In particular, the field redefinition is
\be
A = A_+ + A_-,\qquad
\th = A_+ - A_-.
\ee
Let us denote field strengths of $A_+$ and $A_-$ as
\be
F_+ = dA_+,\qquad
F_- = dA_-.
\ee
Therefore,
\be
F = F_+ + F_-,\qquad
\Th = F_+ - F_-.
\ee
This gives
\be
Y = \cF_+ - \cF_-,\qquad
*Y = -\cF_+ - \cF_-,
\ee
where
\be
\cF_+\equiv F_+ - *F_+,\qquad
\cF_-\equiv F_- + *F_-.
\ee
Substituting these into the Lagrangian \eq{L_Ftheta},
we obtain, up to a total derivative,
\be
\begin{split}
\tilde{\cL}_{\textrm{Maxwell}}
&=-\ove{4}F_+\w*F_+ + \ove{4}\cF_+\w\cP\cF_+
-\ove{4}F_-\w*F_- - \ove{4}\cF_-\w\cP\cF_-\\
&=\cL_{\textrm{PST}}^{(g),(+1)}\big|_{A = A_+}
+\cL_{\textrm{PST}}^{(g),(-1)}\big|_{A = A_-}.
\end{split}
\ee
The Lagrangian $\tilde{\cL}_{\textrm{Maxwell}}$
is separated into two sub-Lagrangians
describing two free chiral 2-forms
of opposite handedness.
Although these sub-Lagrangians share the auxiliary scalar field $a$,
this does not make $A_+$ and $A_-$ couple to each other
since $a$ has no dynamics.

\subsection{A quadratic Sen action is a sum of two PST actions}
\subsubsection{Introducing an extra $2p$-form}

Motivated by the previous subsection,
let us first consider the equation of motion of the Sen formulation.
The equations of motion \eq{EOMP-Sen}-\eq{EOMQ-Sen}
for the quadratic Sen theory are equivalent to
eq.\eq{soln1-Sen}-\eq{soln2-Sen}.
Let us focus on eq.\eq{soln1-Sen},
which is equivalent to
\be\label{Xeq0}
X
\equiv
Q + \hlf d\hat{P} + \hlf\bar{*}dP
\stackrel{on-shell}{=} 0.
\ee

Inspired by eq.\eq{L_Ftheta}, let us consider
the Lagrangian
\be\label{SenX}
\cL
=\ove{4}dP\w\bar{*}dP
-Q\w dP
+\hlf Q\w R
+X\w\bar{\cP}\bar{*}X,
\ee
which, as to be shown below,
has EOM containing eq.\eq{Xeq0}.
The variation of the Lagrangian \eq{SenX} is
\be\label{vary-SenX}
\begin{split}
\d\cL
&=d\d P\w \bar{\cP}X
+d\d\hat{P}\w\bar{\cP}\bar{*}X
+\d Q\w\lrbrk{-\hlf dP + \hlf\bar{*}dP + R + (1-\bar{*})\bar{\cP}\bar{*}X}\\
&\qquad-2d\d a\w\frac{\bar{\cP}\bar{*}X}{(\bar{\pa} a)^2}\w i_{\bar{g}^{-1}da}X.
\end{split}
\ee
The PST1 symmetry is
\be\label{PST1-SenX}
\d P = \chi^{(1)}\w da,\qquad
\d\hat{P} = \chi^{(2)}\w da,\qquad
\d Q = 0 = \d a.
\ee
The PST2 symmetry is
\be
\d P = 2\d a \frac{i_{\bar{g}^{-1}da}\bar{*}X}{(\bar{\pa} a)^2},\qquad
\d \hat{P} = 2\d a \frac{i_{\bar{g}^{-1}da}X}{(\bar{\pa} a)^2},\qquad
\d Q = 0.
\ee

The EOM of $P, \hat{P}, Q, a$ are, respectively,
\be\label{EOMP-SenX}
d\bar{\cP}X = 0,
\ee
\be\label{EOMPh-SenX}
d\bar{\cP}\bar{*}X = 0,
\ee
\be\label{EOMQ-SenX}
-\hlf dP + \hlf\bar{*}dP + R + (1-\bar{*})\bar{\cP}\bar{*}X = 0,
\ee
\be\label{EOMa-SenX}
d\lrbrk{da\w \frac{i_{\bar{g}^{-1}da}\bar{*}X}{(\bar{\pa}a)^2}\w \frac{i_{\bar{g}^{-1}da}X}{(\bar{\pa}a)^2}}
=0.
\ee
Note that the EOM of $a$ follows from the EOM of $P$ and $\hat{P}$.
The equations \eq{EOMP-SenX}-\eq{EOMPh-SenX} are equivalent to
\be
\bar{\cP}\bar{*}X
=d\psi^{(1)}\w da,\qquad
\bar{\cP}X
=d\psi^{(2)}\w da.
\ee
By making use of the PST1 gauge symmetry with
parameters $\chi^{(1)} = 2\psi^{(1)}, \chi^{(2)} = 2\psi^{(2)},$
we obtain
\be
\bar{\cP}X = 0 = \bar{\cP}\bar{*}X,
\ee
and hence
\be\label{X0}
X = 0.
\ee
The only EOM left to be considered is eq.\eq{EOMQ-SenX}.
It becomes, after using eq.\eq{X0},
\be\label{RPP}
-\hlf dP + \hlf\bar{*}dP + R = 0.
\ee
The EOM \eq{EOMP-SenX}-\eq{EOMa-SenX}
are equivalent to eq.\eq{X0}-\eq{RPP},
which in turn are 
equivalent to EOM \eq{EOMP-Sen}-\eq{EOMQ-Sen} of the original Sen theory.

Let us work out diffeomorphism transformation of the theory \eq{SenX}.
Following \cite{Hull:2023dgp}, the Sen action has three types of related diffeomorphisms.
The $\z$-transformation $\d_\z^{\textrm{zeta}}$ transforms $g$ in the standard way
but leaves $\bar{g}$ untransformed.
The $\chi$-transformation $\d_\chi^{\textrm{chi}}$ transforms $\bar{g}$ in the standard way
but leaves $g$ untransformed.
Finally, the standard transformation $\pounds_\xi$
transforms every fields by Lie derivative.
The sum of the $\z$- and $\chi$- transformation
with identified parameter
is equivalent to the standard transformation,
i.e. $\d_\xi^{\textrm{zeta}} + \d_\xi^{\textrm{chi}} = \pounds_\xi$.
Since the action \eq{SenX} is equivalent to the original quadratic Sen action \eq{quadSen},
we may expect that it also has these three types of diffeomorphisms.

In order to work out $\z$-transformation,
it is useful to make use of the identity
\be\label{iden1}
\lrbrk{\frac{1-\bar{*}}{2} + \cM\frac{1+\bar{*}}{2}}\frac{1+*}{2}\l = 0,
\ee
where $\l$ is an arbitrary $(2p+1)$-form.
One may derive the identity \eq{iden1}
by directly working out in coordinate basis
as explained in appendix \ref{app:proofiden}.
Next, applying $\z$-transformation on eq.\eq{iden1} gives
\be
\d_\z^{zeta}\cM\frac{1+\bar{*}}{2}\frac{1+*}{2}
=-\hlf\lrbrk{\frac{1-\bar{*}}{2} + \cM\frac{1+\bar{*}}{2}}(\pounds_\z*).
\ee
Applying the above identity on $H$ gives
\be\label{dzetaM}
(\d_\z^{zeta}\cM) Q
=-\hlf\lrbrk{\frac{1-\bar{*}}{2} + \cM\frac{1+\bar{*}}{2}}(\pounds_\z*)H.
\ee
Applying Lie derivative on eq.\eq{Hsd} gives
\be
(\pounds_\z*)H = (1-*)\pounds_\z H.
\ee
Substituting into eq.\eq{dzetaM}
and using eq.\eq{iden1} gives
\be
\begin{split}
(\d_\z^{zeta}\cM) Q
&=-\lrbrk{\frac{1-\bar{*}}{2} + \cM\frac{1+\bar{*}}{2}}\pounds_\z H.
\end{split}
\ee
So
\be
\begin{split}
Q\w (\d_\z^{zeta}\cM) Q
&=\pounds_\z H\w H\\
&=tot. + 2 di_\z H\w H.
\end{split}
\ee
Suppose that
\be
\d^{zeta}_\z X
=0
=\d^{\textrm{zeta}}_\z a = 0.
\ee
So
\be
\begin{split}
\d^{zeta}_\z \cL
&=tot. +di_\z H\w H
+ d\d^{zeta}_\z P\w\lrbrk{\hlf H +\hlf\bar{\cP}(1-\bar{*})X}\\
&\qquad+d\d^{zeta}_\z \hat{P}\w\lrbrk{\hlf H -\hlf\bar{\cP}(1-\bar{*})X}.
\end{split}
\ee
In order for $\d^{zeta}_\z \cL$ to vanish up to total derivatives,
we set $\d^{\textrm{zeta}}_\z P = -i_{\z}H = \d^{\textrm{zeta}}_\z \hat{P}$.

Summarising, the $\z$-transformation is given by
\be
\d^{\textrm{zeta}}_\z g_{\m\n} = \pounds_{\zeta} g_{\m\n},\qquad
\d^{\textrm{zeta}}_\z \bar{g}_{\m\n} = 0,
\ee
\be
\d^{\textrm{zeta}}_\z P = -i_{\z}H = \d^{\textrm{zeta}}_\z \hat{P},\qquad
\d^{\textrm{zeta}}_\z Q = -\lrbrk{\frac{1+\bar{*}}{2}}\d^{\textrm{zeta}}_\z P,
\ee
\be
\d^{\textrm{zeta}}_\z a = 0.
\ee
The $\chi$-transformation can then be worked out
by using $\d_\chi^{\textrm{chi}} = \pounds_{\chi} - \d_\chi^{\textrm{zeta}}$.
This gives
\be
\d^{\textrm{chi}}_\chi g_{\m\n} = 0,\qquad
\d^{\textrm{chi}}_\chi \bar{g}_{\m\n} = \pounds_{\chi} \bar{g}_{\m\n},
\ee
\be
\d^{\textrm{chi}}_\chi P = \pounds_{\chi}P+i_{\chi}H,\qquad
\d^{\textrm{chi}}_\chi \hat{P} = \pounds_{\chi}\hat{P}+i_{\chi}H,\qquad
\d^{\textrm{chi}}_\chi Q = \pounds_\chi Q-\lrbrk{\frac{1+\bar{*}}{2}}i_\chi H,
\ee
\be
\d^{\textrm{chi}}_\chi a = \pounds_\chi a.
\ee

\subsubsection{A field redefinition and the realisation of separation}

Having shown the equivalence between the Lagrangians \eq{quadSen}
and \eq{SenX} and verifying the symmetries,
let us show the separation at the Lagrangian level.
By making the field redefinition
\be\label{PPAA}
P = A + \bar{A},\qquad
\hat{P} = A - \bar{A},
\ee
we obtain, up to total derivative,
\be\label{LasLALA-pre}
\cL = \cL_{\bar{A}} + \cL_{g-\textrm{Sen}},
\ee
where
\be\label{LAbar}
\cL_{\bar{A}}
=-\hlf d\bar{A}\w\bar{\cP}(1-\bar{*})d\bar{A}
=-\cL^{(\bar{g}),(+1)}_{\textrm{PST}}|_{A\to \bar{A}},
\ee
\be\label{LA-pre}
\begin{split}
\cL_{g-\textrm{Sen}}
&=\hlf dA\w\bar{\cP}(1+\bar{*})dA
-2Q\w(1-\bar{\cP})dA
+Q\w\bar{\cP}Q
+\hlf Q\w R\\
&=\ove{4}dA\w\bar{*}dA - Q\w dA + \hlf Q\w R + H_{(s)}(Q,A)\w\bar{\cP}H_{(s)}(Q,A),
\end{split}
\ee
where $H_{(s)}$ is as defined in eq.\eq{Hsdefn}.
It can be seen that the Lagrangian eq.\eq{LAbar}
is simply the PST Lagrangian with the wrong sign.
As for the Lagrangian \eq{LA-pre},
we will call $g$-Sen Lagrangian
for the reason to be given at the end of subsection \ref{subsec:SenToPST}.

The Lagrangian \eq{LasLALA-pre} can be further worked out.
By using the field redefinition \eq{PPAA}
and noting $(1-\bar{*})\bar{\cP}R = R,$
eq.\eq{EOMQ-SenX} becomes
\be\label{EOMQ-simplified}
(1-\bar{*})(1-\bar{\cP})(dA + H) = 0.
\ee
Acting on the above equation by $(1-\cP)$
and using the identity
\be\label{iden2}
\cP\bar{\cP}\l = \bar{\cP}\l,\qquad
\bar{\cP}\cP\l = \cP\l,
\ee
which holds for any $(2p+1)$-form $\l$,
we obtain
\be\label{piH}
(1-\cP)H = (1-\cP)(-dA).
\ee
Applying $*$ to eq.\eq{piH} and using $H = *H$ gives
\be\label{pH}
\cP H = -\cP*dA.
\ee
Adding eq.\eq{piH}-\eq{pH} together gives
\be
H = -dA + \cP\cF.
\ee
Then by using $Q = (1+\bar{*})H/2$, we obtain
\be\label{Qsoln}
Q
=\lrbrk{\frac{1+\bar{*}}{2}}(-dA + \cP\cF).
\ee
As a cross-check,
we note that the identity \eq{iden1}
implies
\be\label{iden1-1}
\lrbrk{\frac{1-\bar{*}}{2} + \cM\lrbrk{\frac{1+\bar{*}}{2}}}dA
=
\lrbrk{\frac{1-\bar{*}}{2} + \cM\lrbrk{\frac{1+\bar{*}}{2}}}\cP\cF.
\ee
By using the identities \eq{iden2} and \eq{iden1-1},
it can be shown that eq.\eq{EOMQ-simplified} is indeed satisfied
by eq.\eq{Qsoln}.

By substituting eq.\eq{Qsoln} into eq.\eq{LasLALA-pre}, we obtain
\be\label{LasLALA}
\cL = \cL_{\bar{A}} + \cL_A,
\ee
where
\be\label{LA}
\cL_A
= \hlf dA\w\cP(1-*)dA
= \cL^{(g),(+1)}_{\textrm{PST}}.
\ee

The form of eq.\eq{LasLALA} suggests that
the Sen Lagrangian is separated.
It can be seen from the Lagrangians $\cL_{\bar{A}}$ and $\cL_A$, i.e. eq.\eq{LAbar} and eq.\eq{LA},
that $\bar{A}$ is
an unphysical $(+1)$-handed $\bar{g}$-chiral
$(2p)$-form field
whereas $A$ is a physical
$(+1)$-handed $g$-chiral
$(2p)$-form field.

More explicitly, since eq.\eq{LasLALA}
is the sum of two PST Lagrangians (but with $\cL_{\bar{A}}$ having incorrect overall sign),
PST1 and PST2 symmetries as well as EOM
can easily be worked out.
The variation of the Lagrangian \eq{LasLALA}
is, up to total derivative,
\be
\d\cL
=\lrbrk{d\d A-\frac{1}{2} d\d a\w\frac{i_{g^{-1}da}\cF}{(\pa a)^2}}\w\cP\cF
-\lrbrk{d\d \bar{A}-\frac{1}{2} d\d a\w\frac{i_{\bar{g}^{-1}da}\bar{\cF}}{(\bar{\pa} a)^2}}\w\bar{\cP}\bar{\cF},
\ee
where
\be
\cF\equiv dA - *dA,\qquad
\bar{\cF}\equiv d\bar{A} - \bar{*}d\bar{A}.
\ee
The PST1 symmetry is
\be
\d A = -\f da,\qquad
\d\bar{A} = -\bar{\f}da,\qquad
\d a = 0.
\ee
The PST2 symmetry is
\be
\d A = \d a \frac{i_{g^{-1}da}\cF}{(\pa a)^2},\qquad
\d \bar{A} = \d a \frac{i_{\bar{g}^{-1}da}\bar{\cF}}{(\bar{\pa} a)^2}.
\ee
The equations of motion for $A, \bar{A}, a$
are
\be\label{EOMA-SenToPST}
d(\cP\cF) = 0,
\ee
\be\label{EOMAbar-SenToPST}
d(\bar{\cP}\bar{\cF}) = 0,
\ee
\be\label{EOMa-SenToPST}
d(\cP\cF)\w\frac{i_{g^{-1}da}\cF}{(\pa a)^2}
-d(\bar{\cP}\bar{\cF})\w\frac{i_{\bar{g}^{-1}da}\bar{\cF}}{(\bar{\pa} a)^2} = 0.
\ee

It can be seen from the equations of motion that
the unphysical field $\bar{A}$ is decoupled from the physical field $A$.
This is because eq.\eq{EOMA-SenToPST} contains $A$ but not $\bar{A}$,
whereas eq.\eq{EOMAbar-SenToPST} contains $\bar{A}$ but not $A$.
As for eq.\eq{EOMa-SenToPST},
although both $\bar{A}$ and $A$ appear in this equation,
there is actually also no coupling between these fields
since this equation is trivially satisfied
after eq.\eq{EOMA-SenToPST}-\eq{EOMAbar-SenToPST}
are imposed.
In other words, even if $\cL_A$ and $\cL_{\bar{A}}$
share the auxiliary scalar $a$, the fields $A$ and $\bar{A}$
do not couple since $a$ has no dynamics.
The only independent equations of motion are 
eq.\eq{EOMA-SenToPST}-\eq{EOMAbar-SenToPST}.
By using a PST1 symmetry with appropriate gauge parameters,
these equations can be put in the form
\be
\cF = 0,\qquad
\bar{\cF} = 0.
\ee

\subsubsection{Gauge-fixing of PST2 symmetry}

The PST2 symmetry can be used to fix $a$.
In particular, consider fixing $a = t$.
The Lagrangian \eq{LasLALA}
becomes the sum of two HT Lagrangians.
\be\label{SenX-dt}
\cL|_{a = t}
=-\cL_{\textrm{HT}}^{(\bar{g}),(+1)}\bigg|_{A\to\bar{A}}+\cL_{\textrm{HT}}^{(g),(+1)},
\ee
where the HT Lagrangians are defined in eq.\eq{HT-action}.
The equation \eq{SenX-dt}
agrees with one of the results of \cite{Janaun:2024wya}.

In order to recover the modified diffeomorphism
of the gauge-fixed theory, one uses the similar idea
as that given at the end of the subsection \eq{subsec:PST}.
Demanding that $a$ is invariant under the modified diffeomorphism,
which is a sum
of PST2 and standard diffeomorphism,
we obtain
the modified diffeomorphism of $A$ and $\bar{A}$
as
\be
\d_{\xi} A
=\pounds_\xi A
-\xi^0 \frac{i_{g^{-1}dt}\cF}{g^{00}},
\ee
\be
\d_{\xi} \bar{A}
=\pounds_\xi \bar{A}
-\xi^0 \frac{i_{\bar{g}^{-1}dt}\bar{\cF}}{\bar{g}^{00}}.
\ee

\section{A PST action as a quadratic Sen action with constraints}\label{sec:Sen-constraint}

\subsection{A PST action as a $g$-Sen action
}\label{subsec:SenToPST}

In the previous section, we have shown that
the Lagrangian \eq{LasLALA-pre}, i.e. $\cL_{\bar{A}} + \cL_{g-\textrm{Sen}}$
is equivalent to eq.\eq{LasLALA}, i.e. $\cL_{\bar{A}} + \cL_A$.
It is then natural to expect that
the Lagrangian $\cL'_A$ itself is equivalent
to the Lagrangian $\cL_A$.
As to be explicitly shown in this subsection,
this is indeed the case.
Essentially,
the Lagrangian $\cL_{g-\textrm{Sen}}$, eq.\eq{LA-pre},
is the quadratic Sen Lagrangian with
an extra term $H_{(s)}(Q,A)\w\bar{\cP}H_{(s)}(Q,A)$,
which can be thought of as the term which introduces
constraints which remove the degrees of freedom of unphysical $\bar{g}$-chiral form,
leaving only the physical $g$-chiral form.

The variation of the Lagrangian $\cL_{g-\textrm{Sen}}$ is
\be
\begin{split}
	\d\cL_{g-\textrm{Sen}} =
	2\lrbrk{d\d A - d\d a\w\frac{i_{\bar{g}^{-1}da}H_{(s)}}{(\bar{\pa}a)^2}}\w \bar{\cP}H_{(s)}
	+\d Q\w\lrbrk{-dA + R + (1-\bar{*})\bar{\cP}H_{(s)}},
\end{split}
\ee
where, for brevity, we have written $H_{(s)}(Q,A)$ simply as $H_{(s)}$.
It can be seen from the variation that
apart from the tensor gauge symmetry of $A$,
the theory \eq{LA-pre} also has PST1 and PST2 symmetries.
The PST1 symmetry is
\be\label{PST1-SenToPST1}
\d A = -\f\w da,\qquad
\d Q = 0,\qquad
\d a = 0,
\ee
where $\f$ is an arbitrary $(2p-1)$-form.
The PST2 symmetry is
\be\label{PST2-SenToPST1}
\d A = \d a\ \frac{i_{\bar{g}^{-1}da}H_{(s)}}{(\bar{\pa}a)^2},\qquad
\d Q = 0,
\ee
where $\d a$ is arbitrary.

The equations of motion of $Q, A$ and $a$ are
\be\label{QEOM-SenToPST1}
0
=(1-\bar{*})(\bar{\cP} - 1)dA + \lrbrk{\cM + (1-\bar{*})\bar{\cP}}Q,
\ee
\be\label{PEOM-SenToPST1}
d(\bar{\cP}H_{(s)}) = 0,
\ee
\be\label{aEOM-SenToPST1}
d(\bar{\cP}H_{(s)})\w\frac{i_{\bar{g}^{-1}da}H_{(s)}}{(\bar{\pa}a)^2}
	=0.
\ee
From eq.\eq{PEOM-SenToPST1}-\eq{aEOM-SenToPST1},
it is clear that
the equation of motion for $a$
is implied by the equation of motion of $A.$
So $a$ has no dynamics.
We are then left with eq.\eq{QEOM-SenToPST1}-\eq{PEOM-SenToPST1}.
Since eq.\eq{QEOM-SenToPST1} is in fact the same as eq.\eq{EOMQ-simplified},
its solution is given by eq.\eq{Qsoln}.
The solution to eq.\eq{PEOM-SenToPST1} is
\be\label{solnP-SenToPST2}
\bar{\cP}H_{(s)}
=d\th\w da,
\ee
where $\th$ is an arbitrary $(2p-1)$-form.
By applying PST1 symmetry \eq{PST1-SenToPST1},
on eq.\eq{solnP-SenToPST2} and identifying $\f = 2\th$,
we obtain
\be\label{PHs}
\bar{\cP}H_{(s)} = 0.
\ee
By substituting eq.\eq{Qsoln} into eq.\eq{PHs}
and using the identity \eq{iden2},
we obtain
\be
\cP\cF = 0,
\ee
which implies
\be\label{sdsdP}
\cF = 0.
\ee

Therefore EOM for the $g$-Sen theory \eq{LA-pre}
is equivalent to eq.\eq{sdsdP}
which is the $*$-self-duality condition for $dA.$
Furthermore, due to the presence of the non-dynamical scalar field $a$ along with the PST2 symmetry \eq{PST2-SenToPST1},
the $g$-Sen theory is equivalent to the PST theory $\cL_{\textrm{PST}}^{(g),(+1)}$.

By noting that eq.\eq{PHs}
implies
$
H_{(s)} = 0,
$
the $g$-Sen theory can then be thought of as
the quadratic Sen theory with constraints to impose $H_{(s)} = 0$.
This theory is equivalent to the PST theory
which describes a physical $g$-chiral $2p$-form.
In other words, the constraints introduced essentially
eliminate the unphysical $\bar{g}$-chiral $2p$-form
from the original quadratic Sen theory.
This justifies our use of the notation $\cL_{g-\textrm{Sen}}$
and calling it the $g$-Sen Lagrangian.

\subsection{A PST action (with wrong sign) as a $\bar{g}$-Sen action 
}\label{subsec:gbarSen}

A natural question is whether it is possible
to instead eliminate the physical $g$-chiral $2p$-form
from the quadratic Sen theory by introducing appropriate constraints.

It turns out that this is indeed possible
by the introduction of an extra term $\bar{\cP}Q\w H$
to the Sen Lagrangian \eq{quadSen}, along with renaming $P$ to $A$.
More explicitly, consider the Lagrangian
which will be called $\bar{g}$-Sen Lagrangian,
\be\label{Lgbar}
\begin{split}
	\cL_{\bar{g}-\textrm{Sen}}
	&=
	\ove{4}dA\w\bar{*}dA - Q\w dA + \hlf Q\w R +\bar{\cP}Q\w H\\
	&=\ove{4}dA\w\bar{*}dA - Q\w dA - Q\w \bar{\cP}Q,
\end{split}
\ee
where in order to obtain the second equality,
we write $H = Q-R$
and use the identity
\be
Q\w\bar{\cP}R
=Q\w(1-\bar{\cP})R
=\hlf Q\w R.
\ee
The variation of the Lagrangian \eq{Lgbar} is
\be\label{dLgbar}
\begin{split}
	\d \cL_{\bar{g}-\textrm{Sen}}
	&=d\d A\w\lrbrk{\hlf\bar{*}dA + Q}
	+\d Q\w\lrbrk{-dA - 2\bar{\cP} Q}\\
	&\qquad+2d\d a\w\frac{i_{\bar{g}^{-1}da}Q}{(\bar{\pa} a)^2}\w\bar{\cP}Q.
\end{split}
\ee

From the variation, we see that the action \eq{Lgbar}
also has PST1 and PST2 symmetries.
The PST1 symmetry is given by
\be\label{PST1-gbar}
\d Q
=
-\lrbrk{\frac{1+\bar{*}}{2}}(d\f\w da),\qquad
\d A = -\f\w da,\qquad
\d a = 0,
\ee
where $\f$ is an arbitrary $(2p-1)$-form.
The PST2 symmetry is given by
\be
\d A = -\frac{2\d a}{(\bar{\pa}a)^2}i_{\bar{g}^{-1}da}Q,\qquad
\d Q
=-\frac{1+\bar{*}}{2}d\d A.
\ee

The equations of motion of $Q, A,$ and $a$ are
\be\label{EOMQ-barg}
-(1-\bar{*})dA -2(1-\bar{*})\bar{\cP}Q
=0,
\ee
\be\label{EOMP-barg}
d H_{(s)} = 0,
\ee
\be\label{EOMa-barg}
\frac{i_{\bar{g}^{-1}da}Q}{(\bar{\pa} a)^2}\w d\lrbrk{\bar{\cP}Q}
=0.
\ee
The solution to eq.\eq{EOMQ-barg}
is
\be\label{solnQ-barg}
Q = -\lrbrk{\frac{1+\bar{*}}{2}}\bar{\cP}\bar{\cF},
\ee
where $\bar{\cF}\equiv dA - \bar{*}dA.$
Substituting $Q$ from eq.\eq{solnQ-barg}
into eq.\eq{EOMP-barg} gives
\be
0 = d(\bar{\cP}\bar\cF).
\ee
This implies
\be
\bar{\cP}\bar{\cF}
=d\th\w da.
\ee
By using PST1 symmetry \eq{PST1-gbar} with $\f = 2\th,$
we obtain
\be\label{cF0}
\bar{\cF} = 0,
\ee
which is $\bar{*}$-self-duality condition for $A.$
In order to verify that $a$ has no dynamics,
one may eliminate $dA$ from eq.\eq{EOMP-barg}
by using eq.\eq{EOMQ-barg}. This gives
$d(\bar{\cP}Q) = 0$ implying eq.\eq{EOMa-barg}.

By substituting eq.\eq{solnQ-barg}
into eq.\eq{Lgbar}, we obtain the
PST Lagrangian in curved spacetime with metric $\bar{g}.$
More precisely,
\be
\cL_{\bar{g}-\textrm{Sen}}
\stackrel{\eq{solnQ-barg}}{=}-\cL_{\textrm{PST}}^{(\bar{g}),(+1)}.
\ee

By considering eq.\eq{solnQ-barg} and eq.\eq{cF0},
we see that $Q = 0$ on-shell,
which implies that $H = 0$ on-shell.
Therefore, one may interpret the $\bar{g}$-Sen Lagrangian \eq{Lgbar}
as the quadratic Sen Lagrangian with constraints
to impose $H = 0$.

\section{A separation of non-linear Sen actions to non-linear PST actions}\label{sec:Sen-nl}
The results of the sections \ref{sec:Sen-separate}-\ref{sec:Sen-constraint}
suggest the complete relationship between the quadratic Sen theory and the quadratic PST theory
at the Lagrangian and EOM levels.
In this section, we will generalise the discussion by including an external $(2p+1)$-form source
as well as considering the case of non-linear self-duality condition.
Then we will demonstrate explicit calculations in the case of the Sen M5-brane action
by showing how it can be separated into the PST M5-action and an unphysical chiral form action.

\subsection{An analysis on non-linear Sen actions}
The quadratic Sen action can be generalised.
The extension by adding external source
and by non-linearising the self-duality condition
has already been discussed since the original papers by Sen \cite{Sen:2015nph}, \cite{Sen:2019qit}.
Further developments on these extensions
are given for example in \cite{Andriolo:2020ykk}, \cite{Hull:2023dgp}, \cite{Vanichchapongjaroen:2020wza}, \cite{Janaun:2024wya}.

The work by \cite{Janaun:2024wya} considers the Sen actions whose field strength of the
unphysical chiral form is linearly self-dual
while the field strength of the physical chiral form
can be non-linearly self-dual.
In particular, a proof of separation of the two fields
at the Hamiltonian and Lagrangian level is given.
The non-linear Sen Lagrangian is separated into two sub-Lagrangians in the HT formulation.
One of the sub-Lagrangians describes an unphysical $\bar{g}$-chiral form field
with linear self-duality condition without any coupling to other external source.
The other sub-Lagrangian describes a physical $g$-chiral form field
with non-linear self-duality condition and couples to external source.

A $(4p+2)$-dimensional non-linear Sen Lagrangian with external $(2p+1)$-form source $J$
is given by \cite{Vanichchapongjaroen:2020wza}
\be
\cL_{\textrm{nl}}
=\ove{4}dP\w\bar{*}dP - Q\w dP + \hlf Q\w R + \hlf H\w J + U(H^J,g),
\ee
where $H = Q-R, H^J = H+J$. The independent fields of this theory are $P$ and $Q$.
The external fields are $g, \bar{g}, J$.
The composite fields are $R$, $H$, and $H^J$.
The field $R$ is $\bar{*}$-anti-self-dual $R = -\bar{*}R.$
Furthermore, it is defined through the variation
\be
\d_Q\lrbrk{\hlf Q\w R + \hlf H\w J + U(H^J,g)}
=\d Q\w R.
\ee
The field $H^J$ is non-linear self-dual according to
\be\label{nl-HJ}
*H^J = \cV(H^J, g).
\ee
In order for eq.\eq{nl-HJ} to truly be a non-linear self-duality condition,
it should relate half of the components of $H^J$
with the other half.
The question on how to ensure this is not yet fully understood.
Attempts to answer this are present in
\cite{Hull:2023dgp}, \cite{Evnin:2022kqn}, \cite{Janaun:2024wya}.
Therefore, let us simply suppose that $\cV$ in eq.\eq{nl-HJ}
takes a suitable form which makes eq.\eq{nl-HJ} a non-linear self-duality condition.

Consider a Lagrangian
\be\label{Sen-nl}
\tilde{\cL}_{\textrm{nl}}
=\ove{4}dP\w\bar{*}dP - Q\w dP + \hlf Q\w R + \hlf H\w J + U(H^J,g)
+X\w\bar{\cP}\bar{*}X,
\ee
where
$X$ is as defined in eq.\eq{Xeq0},
i.e.
\be
X
\equiv
Q + \hlf d\hat{P} + \hlf\bar{*}dP.
\ee
The Lagrangian $\tilde{\cL}_{\textrm{nl}}$
is equivalent to $\cL_{\textrm{nl}}$.
To see this, we consider the EOM level.
In particular if $X = 0$ on-shell,
then the two theories are equivalent.
The variation of the Lagrangian \eq{Sen-nl}
is exactly given by RHS of eq.\eq{vary-SenX}.
This implies, in particular, the PST1 symmetry takes the form
\eq{PST1-SenX},
while the EOM of $P, \hat{P}, Q, a$
are given by eq.\eq{EOMP-SenX}-\eq{EOMa-SenX}.
Therefore, by following exactly the same procedure to solve these EOM,
we obtain $X = 0$ as required.

Let us eliminate $Q$
by using the equation of motion for the Lagrangian \eq{Sen-nl}.
In particular, the equation of motion for $Q$ is
\be\label{EOMQ-nl}
(1-\bar{*})(\bar{\cP} - 1)(dA + H)
=0,
\ee
where we make use of the field redefinition as in eq.\eq{PPAA}.
The equation \eq{EOMQ-nl}
is the same as the one from the quadratic case,
i.e. eq.\eq{EOMQ-simplified}. The strategy to solve this equation
for $Q$
is therefore similar.
The exception is that we need to apply non-linear self-duality condition
\eq{nl-HJ} instead of the linear one.
Applying $1-\cP$ and using the identity \eq{iden2}, we obtain
\be\label{PiH}
(1-\cP)H^J = -(1-\cP)F,
\ee
where
\be
F \equiv dA - J.
\ee
We then use the
non-linear self-duality condition eq.\eq{nl-HJ}
to express $\cP H^J$ in terms of $g$ and $(1-\cP)H^J$,
the latter of which becomes $-(1-\cP)F$ after eliminating $Q$.
As a result, after $Q$ is eliminated, the expression $U$
is a function of $(1-\cP)F$ and $g$.
In order to eliminate $Q$ from the other terms of the Lagrangian \eq{Sen-nl},
we note that by
applying $1-\bar{\cP}$ to eq.\eq{PiH} and using the
identity \eq{iden2}, we obtain
\be\label{PibarH}
(1-\bar{\cP})H^J = -(1-\bar{\cP})F.
\ee
Then by using
\be
Q = \lrbrk{\frac{1+\bar{*}}{2}}H,
\ee
\be
R = -\lrbrk{\frac{1-\bar{*}}{2}}H,
\ee
along with eq.\eq{PiH} and eq.\eq{PibarH},
we obtain
\be
\begin{split}
	&\ove{4}dP\w\bar{*}dP-Q\w dP + \hlf Q\w R
	+\hlf H^J\w J
	+X\w\bar{\cP}\bar{*}X\\
	&=-\hlf d\bar{A}\w\bar{\cP}(1-\bar{*})d\bar{A}
	+\hlf F\w \cP F+\hlf F\w \cP H^J
			 - \hlf F\w J.
\end{split}
\ee
We may also use eq.\eq{PiH} along with eq.\eq{nl-HJ}
to express $\cP H^J$ in terms of $g$ and $(1-\cP)F$.
Then the expression $F\w\cP H^J = (1-\cP)F\w\cP H^J$
is a function of $(1-\cP)F$ and $g$.
So after eliminating $Q$, we may express
\be
\hlf F\w\cP H^J + U(H^J, g)
=\tilde{U}((1-\cP)F, g).
\ee
The Lagrangian \eq{Sen-nl} then becomes
\be
\tilde{\cL}_{\textrm{nl}}
=-\hlf d\bar{A}\w\bar{\cP}(1-\bar{*})d\bar{A}
	+\hlf F\w \cP F+\tilde{U}((1-\cP)F, g)
			 - \hlf F\w J.
\ee

One may also consider the non-linear $g$-Sen Lagrangian
\be
\cL_{g-\textrm{nl}}
=\cL_{\textrm{nl}}|_{P\to A}
+H_{(s)}(Q,A)\w\bar{\cP}H_{(s)}(Q,A),
\ee
where $H_{(s)}$ is defined in eq.\eq{Hsdefn}.
It can be shown that the unphysical degrees of freedom are eliminated.
The EOM of $Q$ is also given by eq.\eq{EOMQ-nl}.
The field $Q$ can then be eliminated using the similar steps
as above. The upshot is that
the Lagrangian becomes
\be
\tilde{\cL}_{g-\textrm{nl}}
=
\hlf F\w \cP F+\tilde{U}((1-\cP)F, g) - \hlf F\w J.
\ee

\subsection{A PST M5-brane action from A Sen M5-brane action}

Let us apply the discussion in the previous subsection
on the
Sen M5-brane action.
The similar procedure should also apply
for other nonlinear Sen theories of chiral forms.

The Sen M5-brane action \cite{Vanichchapongjaroen:2020wza}
describes an six-dimensional M5-brane worldvolume embedded in
the eleven-dimensional background target space.
The action is written in the Green-Schwarz formalism,
in which the worldvolume theory is not manifestly supersymmetric
while the background supergravity target space is supersymmetric.
The worldvolume action (with $p=1$, i.e. six-dimensional) reads
\be\label{SenM5}
\begin{split}
\cL_{\textrm{Sen-M5}}
&=\ove{4}dP\w\bar{*}dP - Q\w dP
+\ove{24}d^6 x\sqrt{-g}U + \hlf Q\w R
+\hlf H^J\w J + \hlf C_6\\
&\qquad+X\w\bar{\cP}\bar{*}X,
\end{split}
\ee
where
\be
U = -24\sqrt{1+\ove{24}H^J_{\m\n\r}(H^J)^{\m\n\r}},
\ee
while $H^J\equiv H + J$ satisfies nonlinear self-duality condition
\be\label{Hstar-nl}
(*H^J)_{\m\n\r}
=\lrbrk{-\frac{U}{12} + \frac{24}{U}}H^J_{\m\n\r}
+\frac{6}{U}H^J_{\a\b[\m}(H^J)^{\a\b\g}H^J_{\n\r]\g},
\ee
and
$X$ is as defined in eq.\eq{Xeq0}.
We have introduced the last term to the Lagrangian \eq{SenM5}.
Due to the analysis of the previous subsection,
one may verify that this Lagrangian is equivalent
to the original Sen M5-brane Lagrangian derived in \cite{Vanichchapongjaroen:2020wza}.
Note that all the indices appearing explicitly in this section
are raised and lowered by the metric $g$.
The field $R$ is defined from $H = Q-R$, $R = -\bar{*}R$,
and
\be
\d_Q\lrbrk{
\ove{24}d^6 x\sqrt{-g}U + \hlf Q\w R
+\hlf H^J\w J}
=\d Q\w R.
\ee
The fields $J, C_6, g$ are external
and are expressible in terms of pull-backs
of background eleven-dimensional target space superfields.
The field $\bar{g}$ is also external.
After the separation of the Sen M5-brane action,
the field $\bar{g}$ should only couple to the unphysical chiral $2$-form field.

We may simply follow the previous subsection.
The additional explicit calculations to be carried out are to
obtain from eq.\eq{Hstar-nl}
the expression of $\cP H^J$ as a function of $(1-\cP)H^J$
and then substitute into the expression of $U$.

Define
\be
H^J_{\m\n}\equiv H^J_{\m\n\r}v^\r,
\ee
\be
\Ht^J_{\m\n}\equiv \Ht^J_{\m\n\r}v^\r,
\ee
\be
F_{\m\n}\equiv F_{\m\n\r}v^\r,
\ee
\be
\Ft_{\m\n}\equiv \Ft_{\m\n\r}v^\r,
\ee
where
\be
\Ht^J_{\m\n\r}\equiv\frac{\sqrt{-g}}{6}\e_{\m\n\r\a\b\g}(H^J)^{\a\b\g},
\ee
\be
\Ft_{\m\n\r}\equiv\frac{\sqrt{-g}}{6}\e_{\m\n\r\a\b\g}F^{\a\b\g},
\ee
\be
v_\m = \frac{\pa_\m a}{\sqrt{(\pa a)^2}}.
\ee
So eq.\eq{PiH} is equivalent to
\be\label{HtFt}
\Ht^J_{\m\n} = -\Ft_{\m\n}.
\ee
By using eq.\eq{Hstar-nl} and eq.\eq{HtFt},
one obtains
\be\label{Hstar-nl-2}
H^J_{\m\n}
=\frac{-(1+z_1)\Ft_{\m\n} + (\Ft^3)_{\m\n}}{\sqrt{1+z_1 + \frac{z_1^2}{2} - z_2}},
\ee
where
\be
z_1\equiv
\hlf[\Ft^2],\qquad
z_2\equiv
\ove{4}[\Ft^4],
\ee
with $[\cdot]$ being matrix trace with respect to worldvolume index: $[A] = A^\m{}_\m$.
We obtain
\be
U = -12\frac{z_1 + 2}{\sqrt{1+z_1 + \frac{z_1^2}{2} - z_2}}.
\ee
Next, by using eq.\eq{Hstar-nl-2}, we obtain
\be
\ove{24}\sqrt{-g}U d^6 x
\hlf F\w\cP H^J
=-d^6 x\sqrt{-g}\sqrt{1+z_1 + \frac{z_1^2}{2} - z_2}.
\ee
Therefore,
\be\label{SenM5-separate}
\cL_{\textrm{Sen-M5}}
=-\hlf d\bar{A}\w\bar{\cP}(1-\bar{*})d\bar{A}
+\hlf F\w\cP F - d^6 x\sqrt{-g}\sqrt{1+z_1 + \frac{z_1^2}{2} - z_2}
-\hlf F\w J + \hlf C_6.
\ee
The first term of the Lagrangian \eq{SenM5-separate}
is $-\cL_{\textrm{PST}}^{(\bar{g}),+1}$.
The other terms of the Lagrangian \eq{SenM5-separate}
are combined into the PST M5-brane Lagrangian \cite{Bandos:1997ui}.

Instead of introducing the last term $X\w\bar{\cP}\bar{*}X$
of eq.\eq{SenM5}, one may replace this term by $H_{(s)}(Q,P)\w\bar{\cP}H_{(s)}(Q,P)$.
Then based on the analysis given at the end of the last subsection,
the resulting constraint Sen M5-brane Lagrangian
is equivalent, after $Q$ is eliminated, to the PST M5-brane Lagrangian
(without the uncoupled unphysical chiral form).

\section{Conclusion and discussion}\label{sec:conclusion}
In this paper, we consider $(4p+2)$-dimensional
chiral $2p$-form theories in the Sen formulation.
A Sen theory describes two uncoupled chiral form fields
which are chiral with respect to different metrics $g, \bar{g}$.
The field strength of the $g$-chiral (resp. $\bar{g}$-chiral) form
is self-dual with respect to the metric $g$ (resp. $\bar{g}$).
The $\bar{g}$-chiral form field is unphysical due to the wrong sign of the kinetic term,
while the $g$-chiral form is physical.
Independent fields in the Lagrangian of a Sen theory
are $P, Q$, where $Q$ satisfies $Q = \bar{*}Q$.
A certain expression $H$ involving $Q, g, \bar{g}$
is self-dual with respect to the metric $g$.
It is shown in \cite{Evnin:2022kqn}, \cite{Andrianopoli:2022bzr},
that one may eliminate $Q$ at the EOM level.
There is an arbitrary $2p$-form
arises from this elimination process.
Two combinations of $P$ and the arbitrary $2p$-form
are realised, one as a $g$-chiral form
while the other is as a $\bar{g}$-chiral form.
We review this process in eq.\eq{EOMP-Sen}-\eq{soln2-Sen}.
The reference \cite{Janaun:2024wya}
suggests that
after eliminating $Q$ at the Hamiltonian level
and making an appropriate field redefinition in phase space,
the first-order form of the Sen Lagrangian
can be realised as two separate HT Lagrangians.

A natural direction which is the main goal of this paper
is to express a Sen action as a sum of two PST actions
uncoupled from each other. This generalisation from HT to PST case
is non-trivial.
This is because the use of the Hamiltonian analysis in \cite{Janaun:2024wya}
is naturally related to
the HT Lagrangians at the last step
since in both of these considerations,
spacetime coordinates are split into time and space coordinates.
However, it is not clear how
to obtain the PST Lagrangians using similar process.
For example, it is not clear how the field redefinition in phase space
would give rise to the PST scalar $a$.
A better approach, which is carried out
in sections \ref{sec:Sen-separate} and \ref{sec:Sen-nl},
is to work entirely at the Lagrangian level
which makes it more natural to obtain the PST Lagrangians.

At the quadratic level, we consider a quadratic Sen theory \eq{quadSen}.
An extra term is added into this Lagrangian giving the Lagrangian \eq{SenX}.
This extra term
introduces an extra field $\hat{P}$.
To ensure that this does not introduce extra degrees of freedom,
the extra term should provide appropriate constraints.
Direct verification at the EOM level ensures that
EOM of the theory \eq{SenX} is equivalent to EOM of the theory \eq{quadSen}.
Furthermore, by substituting $\hat{P}$ obtained from the EOM
into eq.\eq{SenX}, we obtain eq.\eq{quadSen}. Hence the equivalence at the Lagrangian level is also shown.

We then make a field redefinition as in eq.\eq{PPAA}.
The Lagrangian \eq{SenX} becomes eq.\eq{LasLALA-pre}.
Then solving for $Q$ from the EOM
and substituting into the Lagrangian \eq{LasLALA-pre},
we obtain the Lagrangian \eq{LasLALA}
which is the sum of two PST Lagrangians: $-\cL_{\textrm{PST}}^{(\bar{g}),(+1)}|_{A\to\bar{A}}$
and $\cL_{\textrm{PST}}^{(g),(+1)}$ (see eq.\eq{PST-action} for the definitions).

As shown in section \ref{sec:Sen-nl},
similar steps can be applied to non-linear Sen theories with $(2p+1)$-form source.
As a demonstration, an explicit calculation
is shown that the Sen M5-brane action
is equivalent to a PST M5-brane action
uncoupled from an unphysical quadratic PST action.

Since the Sen formulation describes two chiral forms decoupled from each other,
one may directly impose appropriate constraints
to eliminate one of the chiral forms.
For this, we consider the Lagrangians called the $g$-Sen Lagrangian \eq{LA-pre}
and the $\bar{g}$-Sen Lagrangian \eq{Lgbar}.
After $Q$ is eliminated, the $g$-Sen (resp. $\bar{g}$-Sen) Lagrangian
is equivalent to $\cL_{\textrm{PST}}^{(g),(+1)}$
(resp. $-\cL_{\textrm{PST}}^{(\bar{g}),(+1)}|_{A\to\bar{A}}$).
Note in particular that while the $g$-Sen Lagrangian
couples to both $g$ and $\bar{g}$,
after eliminating $Q$ the coupling with $\bar{g}$ disappears.
This seems to be related to the identity \eq{iden2}.
For example, while the LHS of the second equality of eq.\eq{iden2}
seems to contain both $g$ and $\bar{g},$
the RHS of the same equality only contain $g$ but not $\bar{g}$.

As a summary, the results of this paper show the relationship between
the Sen formulation and the PST formulation
at the Lagrangian level.
Previously, it was a priori unclear how these two formulations
are related at the Lagrangian level.
This is because there is an interchange between
the off-shell and on-shell conditions
between these two formulations.
It was also a priori unclear how the PST scalar $a$
should be introduced in the process
of realising the equivalence of the Sen formulation
with the PST formulation.
It turns out that these issues can be resolved
by combining the following ideas: introducing extra term which contains the auxiliary scalar $a$,
eliminating $Q$, and making an appropriate field redefinition.

It is interesting to extend the result of this paper
to explain the equivalence between the Sen formulation
and the PST formulations with multiple PST scalars \cite{Pasti:2009xc}, \cite{Ko:2013dka}, \cite{Ko:2017tgo}.
This would provide even further insights to the properties of the Sen formulation.

An alternative direction is to see how the Sen formulation
can be realised as being equivalent to the clone formulation \cite{Evnin:2022kqn}, \cite{Mkrtchyan:2019opf}, \cite{Bansal:2023pnr}.

\section*{Acknowledgements}
We would like to thank Sheng-Lan Ko, Anajak Phonchantuek, and Sujiphat Janaun
for discussions.
This research has received funding support from the NSRF via the Program Management Unit for Human Resources \& Institutional Development, Research and Innovation [grant number B39G670016].

\appendix

\section{Proof of the identity \eq{iden1} in coordinate basis}\label{app:proofiden}

In order to derive eq.\eq{iden1}, it is convenient to use
the tools explained in \cite{Janaun:2024wya}.
Let us present crucial set up. For more details, please see \cite{Janaun:2024wya}.

Let an index with angled bracket $\VEV{\cdot}$ stands for the collection of $2p$ indices. For example $\VEV{a} = (a_1\cdots a_{2p}).$ Let an index with square bracket $[\cdot]$ stands for the collection of $2p+1$ indices.
For example $[i] = (i_1\cdots i_{2p+1}),$ $[\m] = (\m_1\cdots\m_{2p+1}).$ Furthermore,
\be
dx^{\VEV{a}}\equiv dx^{a_1}\w\cdots\w dx^{a_{2p}},
\ee
\be
dx^{[i]}\equiv dx^{i_1}\w\cdots\w dx^{i_{2p+1}}.
\ee
Define
\be\label{wMN}
w^{\m_1\cdots\m_{2p+1},\n_1\cdots\n_{2p+1}}
\equiv\sqrt{-\det(\bar{g})}\bar{g}^{[\m_1|\n_1|}\cdots\bar{g}^{\m_{2p+1}]\n_{2p+1}},
\ee
\be\label{vMN}
v^{\m_1\cdots\m_{2p+1},\n_1\cdots\n_{2p+1}}
\equiv\sqrt{-\det(g)}g^{[\m_1|\n_1|}\cdots g^{\m_{2p+1}]\n_{2p+1}},
\ee
where indices $\m_1\cdots\m_{2p+1}$ are totally antisymmetrised and indices $\n_1\cdots\n_{2p+1}$ are also totally antisymmetrised.
So
\be
w^{\m_1\cdots\m_{2p+1},\n_1\cdots\n_{2p+1}}
=w^{\n_1\cdots\n_{2p+1},\m_1\cdots\m_{2p+1}},
\ee
\be
v^{\m_1\cdots\m_{2p+1},\n_1\cdots\n_{2p+1}}
=v^{\n_1\cdots\n_{2p+1},\m_1\cdots\m_{2p+1}}.
\ee

We now use of Dirac bra-ket notation.
We suppress the indices of the form $\VEV{a}$ (which might also include index $0$ when applicable).
Quantities with one set of index $0\VEV{a}$ are represented by ket or bra. In particular,
\be
\begin{split}
	\ket{\e^{[i]}}
	&= \lrbrk{\frac{\e^{0\VEV{a}{[i]}}}{(2p+1)!}} = \bra{\e^{[i]}},\\
	\ket{\e_{[i]}}
	&= \lrbrk{-\frac{\e_{0\VEV{a}{[i]}}}{(2p)!}} = \bra{\e_{[i]}},
\end{split}
\ee
\be\label{ketdTketwi}
\begin{split}
	\ket{dT} &= \lrbrk{dt\w dx^{\VEV{a}}} = \bra{dT},\\
	\ket{w^{[i]}} &= \lrbrk{w^{0\VEV{a},[i]}} = \bra{w^{[i]}}.
\end{split}
\ee
Quantities with two sets of index $0\VEV{a}$ are suppressed
and are considered as linear operators acting on bra or ket.
In particular,
\be
w = (w^{0\VEV{a},0\VEV{b}}),
\ee
\be
v = (v^{0\VEV{a},0\VEV{b}}).
\ee
The contraction of suppressed indices are defined such that one of the indices is upper while the other is lower. For example, the expressions $\ipr{\e^{[i]}}{\e_{[j]}}, \ket{\e^{[i]}}\bra{\e_{[i]}}, w\ket{\e_{[i]}}$ are meaningful. But
the quantity $w\ket{\e^{[i]}}$ is undefined because the suppressed indices are all upper indices, so they cannot be contracted.

Let us denote eigenforms of $\bar{*}$ as $\ket{\bar{e}_{\pm}}$
which corresponds to eigenvalues $\pm 1.$
More precisely,
\be\label{barepm}
\ket{\bar{e}_{\pm}}
\equiv
\ket{dT}+ W_{\pm}\ket{\e_{[i]}}dx^{[i]}.
\ee
It can be worked out that $W_{\pm} = W_{\pm}^T$.
Similarly, we denote 
eigenforms of $*$ as $\ket{e_{\pm}}$
\be\label{epm}
\ket{e_{\pm}}
\equiv
\ket{dT}+ V_{\pm}\ket{\e_{[i]}}dx^{[i]},
\ee
which corresponds to eigenvalues $\pm 1.$
We have $V_{\pm} = V_{\pm}^T$.
In fact, the explicit forms of (the inverses of) $W_{\pm}$ and $V_{\pm}$
are derived in \cite{Janaun:2024wya}.
However, we do not need these forms here.

We can then express
\be
Q = \frac{(-1)^p}{(2p)!}\ipr{Q}{\bar{e}_+},
\ee
\be
R = \frac{(-1)^p}{(2p)!}\ipr{R}{\bar{e}_-},
\ee
\be
H = \frac{(-1)^p}{(2p)!}\ipr{H}{e_+}.
\ee
By using $H = Q - R,$ and
$\cM Q = R,$ we have
\be\label{cMbarep}
\cM\ket{\bar{e}_+}
=(V_+ - W_+)(V_+ - W_-)^{-1}\ket{\bar{e}_-}.
\ee
By using eq.\eq{barepm}-\eq{epm}, \eq{cMbarep},
one obtains
\be
(1-\cM)\lrbrk{\frac{1+\bar{*}}{2}}\lrbrk{\frac{1+*}{2}}\ket{\bar{e}_{\pm}}
=\lrbrk{\frac{1+*}{2}}\ket{\bar{e}_{\pm}},
\ee
and hence
\be
(1-\cM)\lrbrk{\frac{1+\bar{*}}{2}}\lrbrk{\frac{1+*}{2}}
=\lrbrk{\frac{1+*}{2}},
\ee
as required.

\providecommand{\href}[2]{#2}\begingroup\raggedright\endgroup


\begin{thebibliography}{10}

\bibitem{Gross:1984dd}
D.~J. Gross, J.~A. Harvey, E.~J. Martinec, and R.~Rohm, ``{The Heterotic
  String},'' \href{http://dx.doi.org/10.1103/PhysRevLett.54.502}{{\em Phys.
  Rev. Lett.} {\bfseries 54} (1985) 502--505}.

\bibitem{Gueven:1992hh}
R.~Gueven, ``{Black p-brane solutions of D = 11 supergravity theory},''
  \href{http://dx.doi.org/10.1201/9781482268737-16}{{\em Phys. Lett. B}
  {\bfseries 276} (1992) 49--55}.

\bibitem{Green:1981yb}
M.~B. Green and J.~H. Schwarz, ``{Supersymmetrical String Theories},''
  \href{http://dx.doi.org/10.1016/0370-2693(82)91110-8}{{\em Phys. Lett. B}
  {\bfseries 109} (1982) 444--448}.

\bibitem{Siegel:1983es}
W.~Siegel, ``{Manifest Lorentz Invariance Sometimes Requires Nonlinearity},''
  \href{http://dx.doi.org/10.1016/0550-3213(84)90453-X}{{\em Nucl. Phys. B}
  {\bfseries 238} (1984) 307--316}.

\bibitem{McClain:1990sx}
B.~McClain, F.~Yu, and Y.~S. Wu, ``{Covariant quantization of chiral bosons and
  OSp(1,1|2) symmetry},''
  \href{http://dx.doi.org/10.1016/0550-3213(90)90585-2}{{\em Nucl. Phys. B}
  {\bfseries 343} (1990) 689--704}.

\bibitem{Floreanini:1987as}
R.~Floreanini and R.~Jackiw, ``{Selfdual Fields as Charge Density Solitons},''
  \href{http://dx.doi.org/10.1103/PhysRevLett.59.1873}{{\em Phys. Rev. Lett.}
  {\bfseries 59} (1987) 1873}.

\bibitem{Henneaux:1988gg}
M.~Henneaux and C.~Teitelboim, ``{Dynamics of Chiral (Selfdual) $P$ Forms},''
  \href{http://dx.doi.org/10.1016/0370-2693(88)90712-5}{{\em Phys. Lett. B}
  {\bfseries 206} (1988) 650--654}.

\bibitem{Perry:1996mk}
M.~Perry and J.~H. Schwarz, ``{Interacting chiral gauge fields in
  six-dimensions and Born-Infeld theory},''
  \href{http://dx.doi.org/10.1016/S0550-3213(97)00040-0}{{\em Nucl. Phys. B}
  {\bfseries 489} (1997) 47--64},
  \href{http://arxiv.org/abs/hep-th/9611065}{{\ttfamily arXiv:hep-th/9611065}}.

\bibitem{Chen:2010jgb}
W.-M. Chen and P.-M. Ho, ``{Lagrangian Formulations of Self-dual Gauge Theories
  in Diverse Dimensions},''
  \href{http://dx.doi.org/10.1016/j.nuclphysb.2010.04.015}{{\em Nucl. Phys. B}
  {\bfseries 837} (2010) 1--21},
  \href{http://arxiv.org/abs/1001.3608}{{\ttfamily arXiv:1001.3608 [hep-th]}}.

\bibitem{Huang:2011np}
W.-H. Huang, ``{Lagrangian of Self-dual Gauge Fields in Various
  Formulations},''
  \href{http://dx.doi.org/10.1016/j.nuclphysb.2012.03.017}{{\em Nucl. Phys. B}
  {\bfseries 861} (2012) 403--423},
  \href{http://arxiv.org/abs/1111.5118}{{\ttfamily arXiv:1111.5118 [hep-th]}}.

\bibitem{Aganagic:1997zq}
M.~Aganagic, J.~Park, C.~Popescu, and J.~H. Schwarz, ``{World volume action of
  the M theory five-brane},''
  \href{http://dx.doi.org/10.1016/S0550-3213(97)00227-7}{{\em Nucl. Phys. B}
  {\bfseries 496} (1997) 191--214},
  \href{http://arxiv.org/abs/hep-th/9701166}{{\ttfamily arXiv:hep-th/9701166}}.

\bibitem{Ko:2016cpw}
S.-L. Ko and P.~Vanichchapongjaroen, ``{The Dual Formulation of M5-brane
  Action},'' \href{http://dx.doi.org/10.1007/JHEP06(2016)022}{{\em JHEP}
  {\bfseries 06} (2016) 022}, \href{http://arxiv.org/abs/1605.04705}{{\ttfamily
  arXiv:1605.04705 [hep-th]}}.

\bibitem{Pasti:1995ii}
P.~Pasti, D.~P. Sorokin, and M.~Tonin, ``{Note on manifest Lorentz and general
  coordinate invariance in duality symmetric models},''
  \href{http://dx.doi.org/10.1016/0370-2693(95)00463-U}{{\em Phys. Lett. B}
  {\bfseries 352} (1995) 59--63},
  \href{http://arxiv.org/abs/hep-th/9503182}{{\ttfamily arXiv:hep-th/9503182}}.

\bibitem{Pasti:1996vs}
P.~Pasti, D.~P. Sorokin, and M.~Tonin, ``{On Lorentz invariant actions for
  chiral p forms},'' \href{http://dx.doi.org/10.1103/PhysRevD.55.6292}{{\em
  Phys. Rev. D} {\bfseries 55} (1997) 6292--6298},
  \href{http://arxiv.org/abs/hep-th/9611100}{{\ttfamily arXiv:hep-th/9611100}}.

\bibitem{Pasti:1997gx}
P.~Pasti, D.~P. Sorokin, and M.~Tonin, ``{Covariant action for a D = 11
  five-brane with the chiral field},''
  \href{http://dx.doi.org/10.1016/S0370-2693(97)00188-3}{{\em Phys. Lett. B}
  {\bfseries 398} (1997) 41--46},
  \href{http://arxiv.org/abs/hep-th/9701037}{{\ttfamily arXiv:hep-th/9701037}}.

\bibitem{Bandos:1997ui}
I.~A. Bandos, K.~Lechner, A.~Nurmagambetov, P.~Pasti, D.~P. Sorokin, and
  M.~Tonin, ``{Covariant action for the superfive-brane of M theory},''
  \href{http://dx.doi.org/10.1103/PhysRevLett.78.4332}{{\em Phys. Rev. Lett.}
  {\bfseries 78} (1997) 4332--4334},
  \href{http://arxiv.org/abs/hep-th/9701149}{{\ttfamily arXiv:hep-th/9701149}}.

\bibitem{Pasti:2009xc}
P.~Pasti, I.~Samsonov, D.~Sorokin, and M.~Tonin, ``{BLG-motivated Lagrangian
  formulation for the chiral two-form gauge field in D=6 and M5-branes},''
  \href{http://dx.doi.org/10.1103/PhysRevD.80.086008}{{\em Phys. Rev. D}
  {\bfseries 80} (2009) 086008},
  \href{http://arxiv.org/abs/0907.4596}{{\ttfamily arXiv:0907.4596 [hep-th]}}.

\bibitem{Ko:2013dka}
S.-L. Ko, D.~Sorokin, and P.~Vanichchapongjaroen, ``{The M5-brane action
  revisited},'' \href{http://dx.doi.org/10.1007/JHEP11(2013)072}{{\em JHEP}
  {\bfseries 11} (2013) 072}, \href{http://arxiv.org/abs/1308.2231}{{\ttfamily
  arXiv:1308.2231 [hep-th]}}.

\bibitem{Ko:2017tgo}
S.-L. Ko and P.~Vanichchapongjaroen, ``{A covariantisation of M5-brane action
  in dual formulation},'' \href{http://dx.doi.org/10.1007/JHEP01(2018)072}{{\em
  JHEP} {\bfseries 01} (2018) 072},
  \href{http://arxiv.org/abs/1712.06408}{{\ttfamily arXiv:1712.06408
  [hep-th]}}.

\bibitem{Sen:2015nph}
A.~Sen, ``{Covariant Action for Type IIB Supergravity},''
  \href{http://dx.doi.org/10.1007/JHEP07(2016)017}{{\em JHEP} {\bfseries 07}
  (2016) 017}, \href{http://arxiv.org/abs/1511.08220}{{\ttfamily
  arXiv:1511.08220 [hep-th]}}.

\bibitem{Sen:2019qit}
A.~Sen, ``{Self-dual forms: Action, Hamiltonian and Compactification},''
  \href{http://dx.doi.org/10.1088/1751-8121/ab5423}{{\em J. Phys. A} {\bfseries
  53} no.~8, (2020) 084002}, \href{http://arxiv.org/abs/1903.12196}{{\ttfamily
  arXiv:1903.12196 [hep-th]}}.

\bibitem{Sen:2015uaa}
A.~Sen, ``{BV Master Action for Heterotic and Type II String Field Theories},''
  \href{http://dx.doi.org/10.1007/JHEP02(2016)087}{{\em JHEP} {\bfseries 02}
  (2016) 087}, \href{http://arxiv.org/abs/1508.05387}{{\ttfamily
  arXiv:1508.05387 [hep-th]}}.

\bibitem{Andriolo:2020ykk}
E.~Andriolo, N.~Lambert, and C.~Papageorgakis, ``{Geometrical Aspects of An
  Abelian (2,0) Action},''
  \href{http://dx.doi.org/10.1007/JHEP04(2020)200}{{\em JHEP} {\bfseries 04}
  (2020) 200}, \href{http://arxiv.org/abs/2003.10567}{{\ttfamily
  arXiv:2003.10567 [hep-th]}}.

\bibitem{Hull:2023dgp}
C.~M. Hull, ``{Covariant action for self-dual p-form gauge fields in general
  spacetimes},'' \href{http://dx.doi.org/10.1007/JHEP04(2024)011}{{\em JHEP}
  {\bfseries 04} (2024) 011}, \href{http://arxiv.org/abs/2307.04748}{{\ttfamily
  arXiv:2307.04748 [hep-th]}}.

\bibitem{Vanichchapongjaroen:2020wza}
P.~Vanichchapongjaroen, ``{Covariant M5-brane action with self-dual 3-form},''
  \href{http://dx.doi.org/10.1007/JHEP05(2021)039}{{\em JHEP} {\bfseries 05}
  (2021) 039}, \href{http://arxiv.org/abs/2011.14384}{{\ttfamily
  arXiv:2011.14384 [hep-th]}}.

\bibitem{Andriolo:2021gen}
E.~Andriolo, N.~Lambert, T.~Orchard, and C.~Papageorgakis, ``{A path integral
  for the chiral-form partition function},''
  \href{http://dx.doi.org/10.1007/JHEP04(2022)115}{{\em JHEP} {\bfseries 04}
  (2022) 115}, \href{http://arxiv.org/abs/2112.00040}{{\ttfamily
  arXiv:2112.00040 [hep-th]}}.

\bibitem{Evnin:2022kqn}
O.~Evnin and K.~Mkrtchyan, ``{Three approaches to chiral form interactions},''
  \href{http://dx.doi.org/10.1016/j.difgeo.2023.102016}{{\em Differ. Geom.
  Appl.} {\bfseries 89} (2023) 102016},
  \href{http://arxiv.org/abs/2207.01767}{{\ttfamily arXiv:2207.01767
  [hep-th]}}.

\bibitem{Phonchantuek:2023iao}
A.~Phonchantuek and P.~Vanichchapongjaroen, ``{Double dimensional reduction of
  M5-brane action in Sen formalism},''
  \href{http://dx.doi.org/10.1140/epjc/s10052-023-11892-2}{{\em Eur. Phys. J.
  C} {\bfseries 83} no.~8, (2023) 721},
  \href{http://arxiv.org/abs/2305.04861}{{\ttfamily arXiv:2305.04861
  [hep-th]}}.

\bibitem{Janaun:2024wya}
S.~Janaun, A.~Phonchantuek, and P.~Vanichchapongjaroen, ``{Nonlinear chiral
  forms in the Sen formulation},''
  \href{http://dx.doi.org/10.1140/epjc/s10052-024-13207-5}{{\em Eur. Phys. J.
  C} {\bfseries 84} no.~8, (2024) 832},
  \href{http://arxiv.org/abs/2404.05380}{{\ttfamily arXiv:2404.05380
  [hep-th]}}.

\bibitem{Pasti:1996va}
P.~Pasti, D.~P. Sorokin, and M.~Tonin, ``{Comment on `Covariant duality
  symmetric actions'},'' \href{http://dx.doi.org/10.1103/PhysRevD.56.2473}{{\em
  Phys. Rev. D} {\bfseries 56} (1997) 2473--2474},
  \href{http://arxiv.org/abs/hep-th/9607171}{{\ttfamily arXiv:hep-th/9607171}}.

\bibitem{Bandos:2014bva}
I.~Bandos, ``{On Lagrangian approach to self-dual gauge fields in spacetime of
  nontrivial topology},'' \href{http://dx.doi.org/10.1007/JHEP08(2014)048}{{\em
  JHEP} {\bfseries 08} (2014) 048},
  \href{http://arxiv.org/abs/1406.5185}{{\ttfamily arXiv:1406.5185 [hep-th]}}.

\bibitem{Tseytlin:1990ar}
A.~A. Tseytlin and P.~C. West, ``{TWO REMARKS ON CHIRAL SCALARS},''
  \href{http://dx.doi.org/10.1103/PhysRevLett.65.541}{{\em Phys. Rev. Lett.}
  {\bfseries 65} (1990) 541--542}.

\bibitem{Sonnenschein:1988ug}
J.~Sonnenschein, ``{CHIRAL BOSONS},''
  \href{http://dx.doi.org/10.1016/0550-3213(88)90339-2}{{\em Nucl. Phys. B}
  {\bfseries 309} (1988) 752--770}.

\bibitem{Andrianopoli:2022bzr}
L.~Andrianopoli, C.~A. Cremonini, R.~D'Auria, P.~A. Grassi, R.~Matrecano,
  R.~Noris, L.~Ravera, and M.~Trigiante, ``{M5-brane in the superspace
  approach},'' \href{http://dx.doi.org/10.1103/PhysRevD.106.026010}{{\em Phys.
  Rev. D} {\bfseries 106} no.~2, (2022) 026010},
  \href{http://arxiv.org/abs/2206.06388}{{\ttfamily arXiv:2206.06388
  [hep-th]}}.

\bibitem{Mkrtchyan:2019opf}
K.~Mkrtchyan, ``{On Covariant Actions for Chiral $p-$Forms},''
  \href{http://dx.doi.org/10.1007/JHEP12(2019)076}{{\em JHEP} {\bfseries 12}
  (2019) 076}, \href{http://arxiv.org/abs/1908.01789}{{\ttfamily
  arXiv:1908.01789 [hep-th]}}.

\bibitem{Bansal:2023pnr}
S.~Bansal, ``{Manifestly covariant polynomial M5-brane lagrangians},''
  \href{http://dx.doi.org/10.1007/JHEP01(2024)087}{{\em JHEP} {\bfseries 01}
  (2024) 087}, \href{http://arxiv.org/abs/2307.13449}{{\ttfamily
  arXiv:2307.13449 [hep-th]}}.

\end{thebibliography}
\end{document}